\documentclass[%
 reprint,
 amsmath,amssymb,
 aps,
]{revtex4-2}

\usepackage{graphicx}
\usepackage{dcolumn}
\usepackage{bm}
\usepackage{bbm}
\usepackage{subcaption}
\usepackage{comment}
\usepackage{physics}
\usepackage{amsmath}

\begin{document}

\preprint{APS/123-QED}

\title{Simple way to incorporate loss when modelling multimode entangled state generation}
\author{Colin Vendromin}
\email{colin.vendromin@queensu.ca}
\author{Marc M. Dignam}
\affiliation{%
Department of Physics, Engineering Physics, and Astronomy, Queen's University, Kingston, Ontario K7L 3N6, Canada
}%

\date{\today}
\begin{abstract}
 We show that the light generated via spontaneous four-wave mixing or parametric down conversion
in multiple, coupled, lossy cavities is a multimode squeezed thermal state. Requiring this state to be the solution of the Lindblad master equation results in a set of coupled first-order differential equations for the time-dependent squeezing parameters and thermal photon numbers of the state. The benefit of this semi-analytic approach is that the number of coupled equations scales linearly with the number of modes and is independent of the number of photons generated. With this analytic form of the state, correlation variances are easily expressed as analytic functions of the time-dependent mode parameters. Thus, our solution makes it computationally tractable and relatively straight forward to calculate the generation and evolution of multimode entangled states in multiple coupled, lossy cavities, even when there are a large number of modes and/or photons.
\end{abstract}

\maketitle

\section{Introduction}
\begin{figure}[h!]
\captionsetup{justification=raggedright}
     \centering
     \begin{subfigure}[b]{0.23\textwidth}
         \centering
         \includegraphics[width=\textwidth]{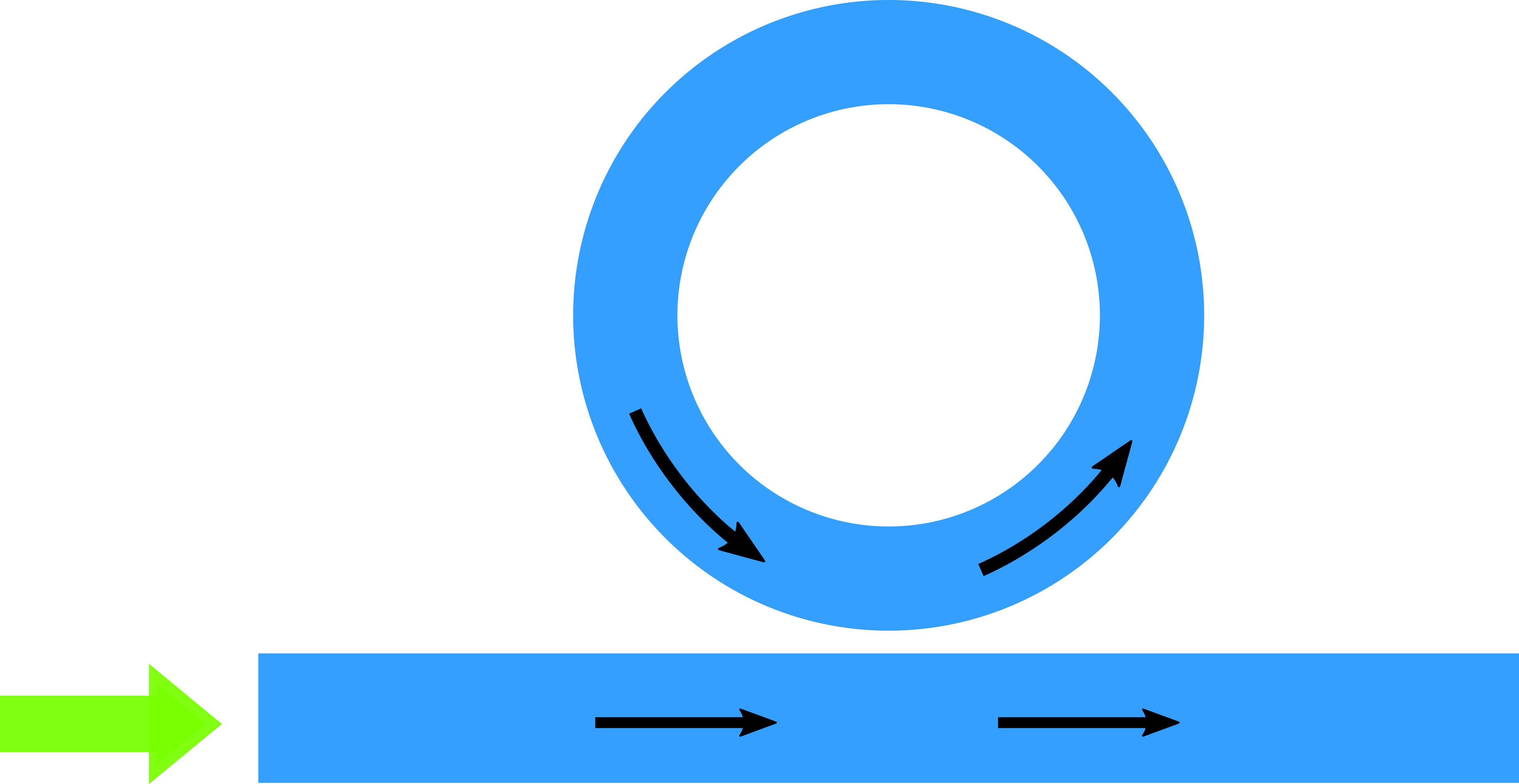}
         \caption{}
         \label{fig:ring}
     \end{subfigure}
     \hfill
     \begin{subfigure}[b]{0.24\textwidth}
         \centering
         \includegraphics[width=\textwidth]{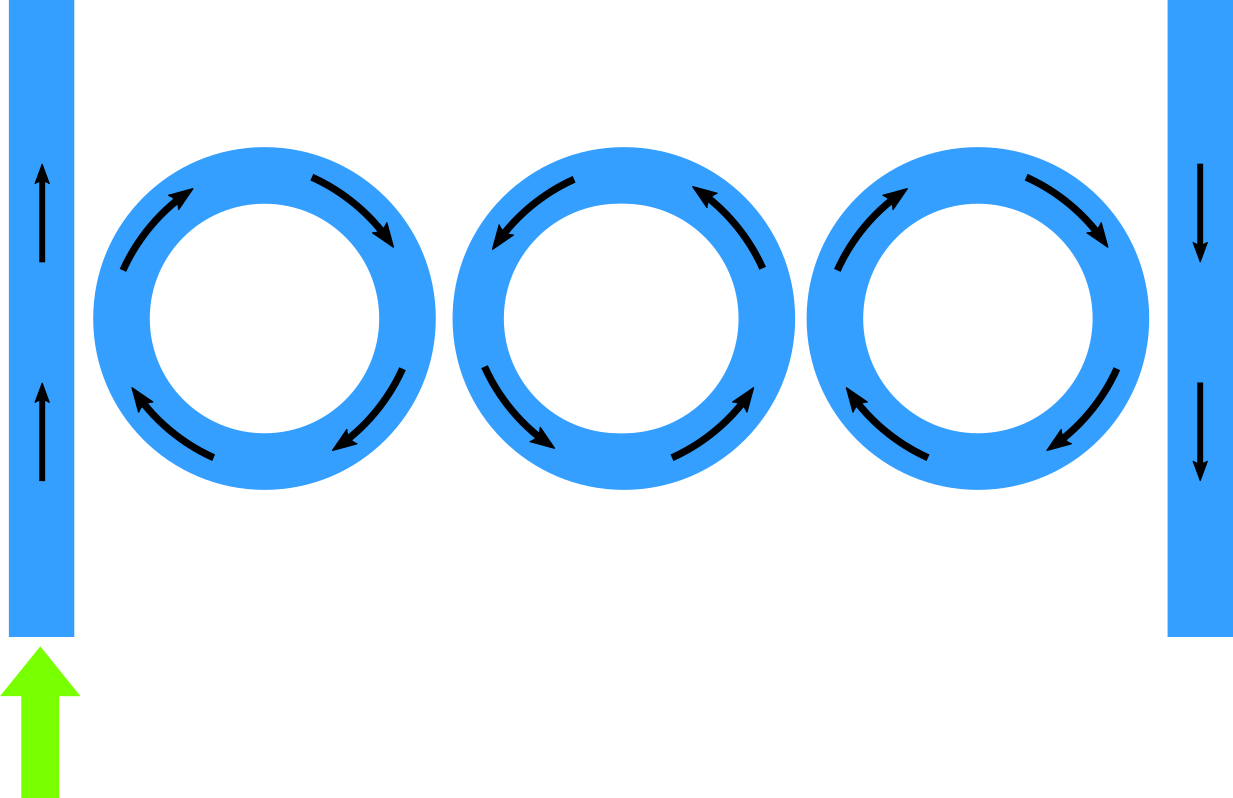}
         \caption{}
         \label{fig:coupledrings}
     \end{subfigure}
     \hfill
     \begin{subfigure}[b]{0.23\textwidth}
         \centering
    \includegraphics[width=\textwidth]{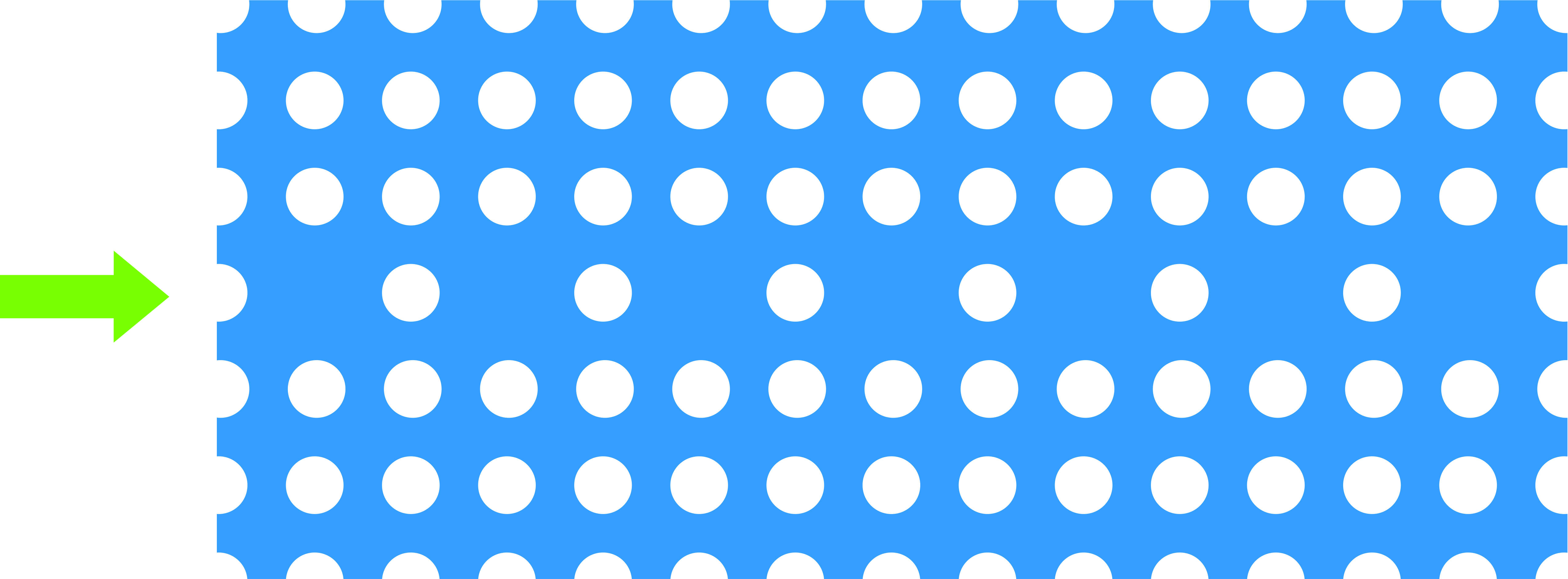}
         \caption{}
         \label{fig:CROW}
     \end{subfigure}
     \hfill
     \begin{subfigure}[b]{0.23\textwidth}
         \centering
                  \includegraphics[width=\textwidth]{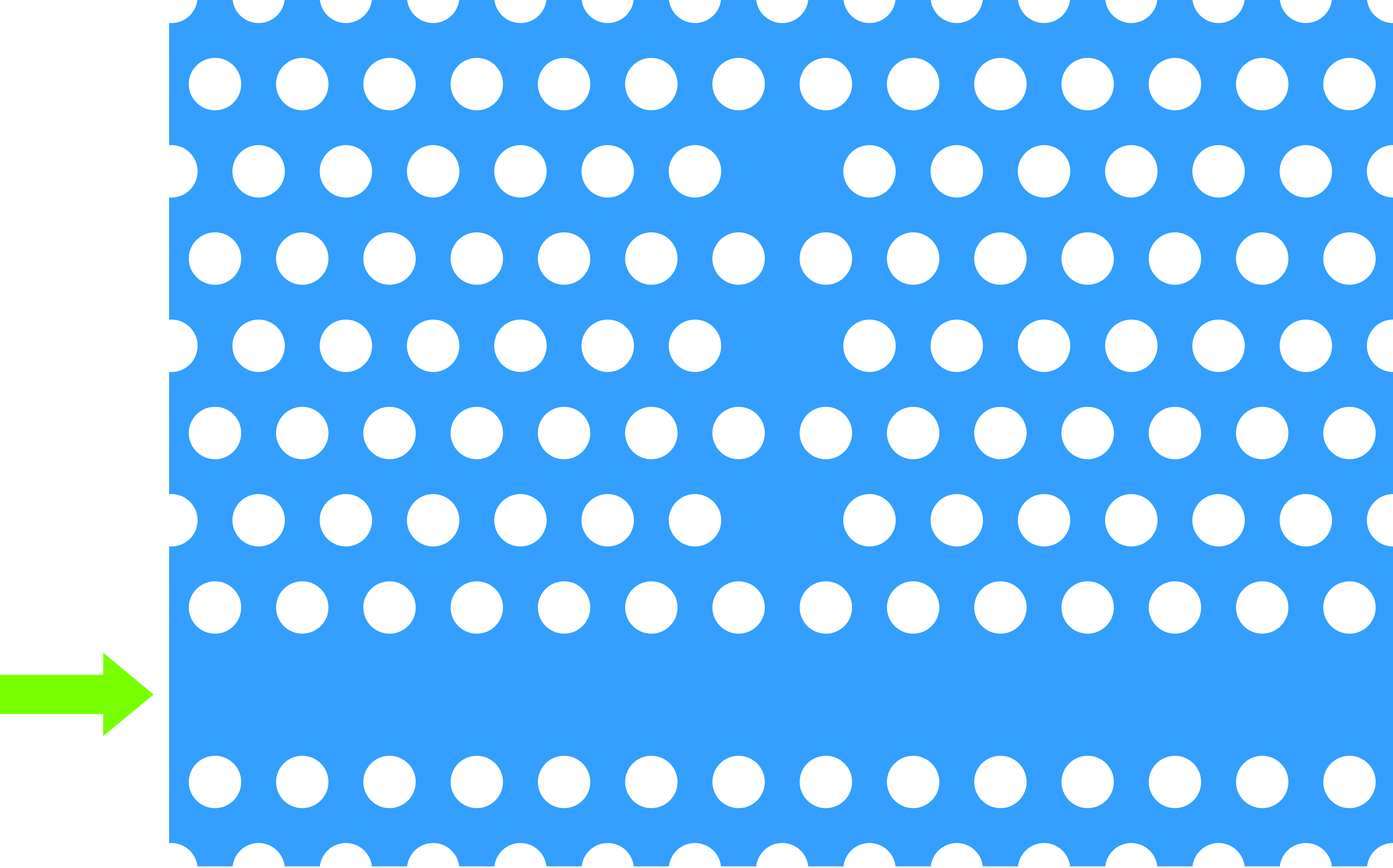}
         \caption{}
         \label{fig:CoupledCavities}
     
     \end{subfigure}
          \caption{Schematics of structures of potential interest. The thick arrow indicates where the pump field is injected. The first structure (a) is a ring resonator coupled to a waveguide. The second (b) are a series of coupled ring resonators coupled to an input and output waveguide. The thin black arrows in (a) and (b) indicate the direction of the field components at the coupling points. The third (c) is a CROW made by periodically removing air holes (shown as white circles) along a row in the crystal. And the fourth (d) is three high-Q cavities that are coupled to a waveguide.}
          \label{fig:structures}
\end{figure}

Multimode squeezed states can be generated via a nonlinear interaction in resonant structures, such as ring resonators (see Fig. \ref{fig:ring}) or coupled-resonator optical waveguides (CROWs) in a photonic crystal (see Fig. \ref{fig:CROW}). They are a source of continuous-variable (CV) entanglement, since the quadratures of the photons in different modes in the state can be correlated. CV entanglement has applications in boson sampling \cite{GaussianBosonSampling, DetailedGaussianBosonSampling}, quantum computing \cite{Braunstein2005QuantumVariables, Takeda2019TowardComputing, CVquantumComputing}, and CV cluster states \cite{CVclusterstateBraunstein, graphicalCalcGuassianStates, wuCVchip2020}.

The theoretical generation of multimode squeezed states via spontaneous four-wave mixing (SFWM) or spontaneous parametric down conversion (SPDC) has been studied extensively for ring resonators \cite{strongquantumopticsringresonator}, nonlinear waveguides \cite{backwardHeisenberg,brightsqueezedvacuum}, and CROWs \cite{Seifoory2019counterpropagatingCV,pairGenerationCrow}. Photon loss is an important problem in these systems, since it can reduce the squeezing and inseparability of the state. 

Loss in ring resonators and waveguides due to photon scattering  can be handled by introducing  reservoir modes that photons can couple in to \cite{strongquantumopticsringresonator,degenerateSqueezingUnified}. The waveguide-reservoir coupling parameters can be estimated with phenomenological values taken from experiment. In contrast, loss in CROWs or coupled-cavities (see Fig. \ref{fig:CoupledCavities}) can be handled intrinsically by calculating the complex frequencies of the CROW Bloch modes or cavity modes \cite{Seifoory2019counterpropagatingCV, pairGenerationCrow}. In this approach, the evolution in these lossy systems is expressed as the  non-unitary evolution of the reduced density operator of the generated light, obtained from the solution of the Lindblad master equation. This is the approach that we use in this work.

Before discussing our approach, we present in Fig. \ref{fig:structures}, four different lossy coupled-cavity structures in which our theoretical method can be used to determine the quantum state of light generated via SFWM or SPDC. In Fig. \ref{fig:ring},  a ring resonator coupled to a waveguide is shown.  The  pump pulse is injected into the straight waveguide; it then couples into the ring, where it generates multimode entangled light via  SFWM or SPDC in the multiple, lossy modes of the ring. In Fig. \ref{fig:coupledrings} a series of three coupled rings is shown. This structure allows for the generation and propagation of the multimode state in the coupled-rings. Fig. \ref{fig:CROW} shows a CROW in a photonic crystal.  The pump pulse propagates in a set of Bloch modes of the CROW and generates signal and idler photons as it propagates. Lastly, Fig. \ref{fig:CoupledCavities} shows a three-mode cavity coupled to a waveguide in a photonic crystal. The state is predominantly generated inside the cavity and couples out into the waveguide.

In this work, we prove that  the density  operator for entangled light generated in multiple lossy modes via SFWM or SPDC takes  the  form  of  a  multimode  squeezed thermal state (MSTS), where the thermal part of the state captures the effects of photon loss. The loss can be handled either phenomenologically or intrinsically, depending on the structure. For $M$ modes, we derive a set of $3M$ coupled first-order differential equations that provide the complete evolution of the squeezing parameters and thermal photon numbers in the density operator. The number of equations is independent of the number of generated photons which is generally many fewer equations than must be solved when using numerical methods  that rely on Fock states \cite{numericalLindblad}, since in those methods the number of equations can be as large as $(M+N)![M!N!]^{-1}$, where $N$ is the total photon number. 

Our solution of the Lindblad master equation applies to the discrete modes of a structure; for example, the modes of the rings in Figs. \ref{fig:ring} and \ref{fig:coupledrings}, and the Bloch modes of the CROW in Fig. \ref{fig:CROW}. Our solution can also be applied to the structure in Fig. \ref{fig:CoupledCavities} as long as the modes are discretized (\textit{e.g.} by using periodic boundary conditions).

In the limit that the modes are lossless, our formalism reproduces the results for a multimode squeezed vacuum state given by N. Quesada \textit{et al.} \cite{beyondPhotonPairs}. Also, in the case of only one  or two  lossy modes, our multimode formalism reproduces the results of previous work \cite{squeezedthermal,Seifoory2017SqueezedCavities,twomodeSL}.

The rest of the paper is organized as follows. In Sec. \ref{quasimodes} we define the lossy modes of a structure, called quasimodes. In Sec. \ref{SFWMquasimodes} we present the nonlinear Hamiltonian for SFWM, and the Lindblad master equation in the basis of quasimodes.  In Sec. \ref{solution} we show that the analytic solution to the Lindblad master equation is the density operator for a multimode squeezed thermal state, and obtain a set of coupled first-order differential equations for the squeezing amplitudes, squeezing phases, and thermal photons numbers. In Sec. \ref{expectation} we give expressions for the correlation variance and the expectation value of the photon number operator. In Sec. \ref{lmitingcases} we show that in the limit when the modes are lossless  our coupled equations describe a squeezed vacuum state that agrees with other work, and that they produce results that are consistent with previous work on single- and two- mode squeezed states. In Sec. \ref{results} we present our results from solving the coupled differential equations for a squeezing process in the Bloch modes of a four-cavity CROW. Finally, in Sec. \ref{Conclusion} we conclude. Additionally, there are three appendices. In Appendix \ref{SVD+takagi} we derive the Takagi factorization, starting from the symmetric singular value decomposition. In Appendix \ref{appndxsolvingeqs} we give the details on deriving the coupled equations for the squeezing process, and in Appendix \ref{solving} we discuss how to numerically solve them for a general system initially in the vacuum state.

\section{Quasimodes}
\label{quasimodes}
The positive frequency part of the electric field in a lossy mode $m$ takes the form 
\begin{equation}
    \label{eq:Efield}
    \tilde{\boldsymbol{E}}^{(+)}_m(\boldsymbol{r},t) = \tilde{\boldsymbol{N}}_m(\boldsymbol{r}){\rm e}^{-i\tilde{\omega}_mt},
\end{equation}
where $\tilde{\boldsymbol{N}}_m(\boldsymbol{r})$ is the spatial profile of the mode and we define $\tilde{\omega}_m = \omega_m - i\gamma_m$  the complex frequency with real part $\omega_m$, and the imaginary part $\gamma_m$ quantifies the energy leakage.  The spatial profile solves the Helmholtz equation
\begin{align}
    \label{eq:helmholtz}
    \boldsymbol{\nabla}\times \boldsymbol{\nabla} \times \tilde{\boldsymbol{N}}_m(\boldsymbol{r}) - \frac{\tilde{\omega}^2_m}{c^2}\epsilon(\boldsymbol{r}) \tilde{\boldsymbol{N}}_m(\boldsymbol{r})&=0,
\end{align}
where $\epsilon(\boldsymbol{r})$ is the relative dielectric function of the structure. Throughout this paper we refer to the solutions of Eq. \eqref{eq:helmholtz} as quasimodes \cite{quasimodeapproach}, and we assume the quasimodes are orthonormal in the following sense
\begin{align}
    \label{eq:orthogonality}
    \int d^3 \boldsymbol{r} \epsilon(\boldsymbol{r}) \tilde{\boldsymbol{N}}^*_m(\boldsymbol{r})\cdot \tilde{\boldsymbol{N}}_l(\boldsymbol{r})&=\delta_{ml}.
\end{align}
The orthogonality of the modes is strictly obeyed if there is symmetry in the structure (\textit{e.g.} rotational or translational). For example, the Bloch modes of the CROW in Fig. \ref{fig:CROW} are orthogonal due to the translational symmetry of the CROW, and the modes of the ring resonator are orthogonal due to the rotational symmetry of the ring. In structures that lack symmetry, such as the cavity coupled to a waveguide in Fig. \ref{fig:CoupledCavities}, the orthonormality in Eq. \eqref{eq:orthogonality} is not generally strictly obeyed. However, modes that are separated in frequency by more than their linewidths will be very nearly orthogonal \cite{spontaneousEmissionSuppression}.  Therefore, in what follows, we will assume mode orthogonality.

In the next section we present a quantized theory of spontaneous four-wave mixing and the Lindblad master equation in the basis of these quasimodes. 

\section{The Lindblad master equation for a multimode lossy structure}
\label{SFWMquasimodes}
Before discussing the solution to the Lindblad master equation for multiple lossy modes, we present the system Hamiltonian in the undepleted pump approximation and the Lindblad master equation that we reference throughout the paper. 

In a lossy dielectric structure, a discrete basis of orthogonal quasimodes can be used to construct the system Hamiltonian
\begin{align}
    \label{eq:systemH}
   H = H_{\rm L}+H_{\rm NL},
\end{align}
where linear Hamiltonian in the basis of the quasimodes is
\begin{equation}
    \label{eq:HL}
    H_{\rm L} = \sum_m\hbar \tilde{\omega}_m b^\dagger_m b_m,
\end{equation}
where $b^\dagger_m$ and $b_m$ are the creation and annihilation operators for photons in the $m{\rm th}$ quasimode satisfying the standard commutation relation
\begin{equation}
    \label{eq:bcommutator}
    \left[b_m,b^\dagger_{l}\right]=\delta_{ml}.
\end{equation}

Now we include the  SFWM nonlinear interaction where two pump photons $p_1$ and $p_2$ are annihilated to create a pair of signal and idler photons $m$ and $l$. The nonlinear Hamiltonian in the basis of the quasimodes is \cite{twinbeamwaveguide}
\begin{equation}
    \label{eq:HNL}
    H_{\rm NL} = \sum_{m,l,p_1,p_2 }G_{mlp_1p_2}b^\dagger_{m} b^\dagger_{l}b_{p_1} b_{p_2}  + {\rm H.c.},
\end{equation}
where
\begin{align}
    \label{eq:nonlinearparameter}
    G_{m l p_1 p_2} &\equiv \frac{9\hbar^2}{16\epsilon_0}\sqrt{\omega_{m}\omega_{l}\omega_{p_1}\omega_{p_2}}\sum_{i,j,k,h}\int d^3 \boldsymbol{r} \chi^{(3)}_{ijkh}(\boldsymbol{r})\nonumber
    \\
    &\times\tilde{N}^*_{m i}(\boldsymbol{r})\tilde{N}^*_{l j}(\boldsymbol{r})\tilde{N}_{p_1 k}(\boldsymbol{r})\tilde{N}_{p_2 h}(\boldsymbol{r})
\end{align}
is the effective nonlinear parameter that depends on the spatial overlap of the quasimodes $\tilde{\boldsymbol{N}}_m(\boldsymbol{r})$. The subscripts $i,j,k$, and $h$ label the Cartesian components of the electric field and the medium nonlinear tensor $\chi^{(3) }$. In deriving this result we have assumed that the imaginary part of the frequencies are small, such that we can neglect them under the square root, so that  $\sqrt{\tilde{\omega}_m}\approx \sqrt{\omega_m}$. This is valid for modes that have a quality factor on the order of $10^3$ or higher. For a CROW studied in previous work \cite{engineeringQ}, it was shown that the quality factors of the Bloch modes varied from approximately $10^4$ to $10^3$. The relative dielectric function does not appear in $H_{\rm NL}$, because we have constructed it using the electric fields. If we used the displacement fields, then the relative dielectric function would appear in the nonlinear parameter in Eq. \eqref{eq:nonlinearparameter}.

The dynamics of the density operator $\rho$ for the pump, signal, and idler light can be modeled using the Lindblad master equation \cite{openQsystemsBreuer}
\begin{align}
    \label{eq:lindblad}
    \frac{d\rho}{dt} = -\frac{i}{\hbar}\left[H,\rho\right] + \sum_m \gamma_m\left(2b_m\rho b^\dagger_m - b^\dagger_m b_m \rho - \rho b^\dagger_m b_m\right),
\end{align}
where $H$ is the Hermitian part of the Hamiltonian in Eq. \eqref{eq:systemH}, given by
\begin{align}
\label{eq:herH}
    H&=\sum_m\hbar\omega_m b^\dagger_m b_m +\sum_{m,l,p_1,p_2 }G_{mlp_1p_2}b^\dagger_{m} b^\dagger_{l}b_{p_1} b_{p_2}  + {\rm H.c.}.
\end{align}
General expressions for the Hamiltonian and Lindblad master equation, that are applicable to any lossy structure, can be derived using a set of non-orthogonal quasimodes \cite{photonQuantumDot}. The extension from the orthogonal to the non-orthogonal case is not straightforward. All derivations, however, in this paper are done assuming the quasimodes are orthogonal according to Eq. \eqref{eq:orthogonality}. 

\subsection{Restrictions on the pump field and nonlinear parameter}
\label{pumpconsiderations}
The Hamiltonian presented in Eq. \eqref{eq:herH} is valid for a general pump field with a quantum description.  However, since we are concerned with generating squeezed states, we let the pump be a classical field and  make the undepleted pump approximation. Furthermore, we restrict our analysis to pump fields that can be factored into a function of space multiplied by a function of time. These two restrictions on the pump are shown in Sec. \ref{solution} to be necessary in order to derive our solution to the Lindblad master equation.

To make the undepleted pump approximation, we let the classical pump field be represented by a lossy multimode coherent state $\ket{\alpha(t)}$, defined as a product of single-mode coherent states $\ket{\alpha_p(t)}$ 
\begin{align}
    \label{eq:multicoherentstate}
    \ket{\alpha(t)} &\equiv \prod_p \ket{\alpha_p(t)},
\end{align}
where the single-mode coherent states are defined as
\begin{align}
    \label{eq:singlecoherentstate}
    \ket{\alpha_p(t)}&= \exp(\alpha_p {\rm e}^{-i\tilde{\omega}_p t} b^\dagger_p -{\rm H.c.})\ket{{\rm vac}}.
\end{align}
For a mode $p$ with a frequency in the pump bandwidth we have
\begin{align}
    \label{eq:singlecoherentstate_operate}
    b_p\ket{\alpha(t)}&= \alpha_p {\rm e}^{-i\tilde{\omega}_p t}\ket{\alpha(t)},
\end{align}
where $\alpha_p$ is the pump amplitude in the $p$th mode.
The average total photon number for the pump is 
\begin{align}
    \label{eq:pumpphotons}
   \bra{\alpha(t)}\sum_p b^\dagger_p b_p\ket{\alpha(t)}=\sum_p |\alpha_p|^2{\rm e}^{-2 \gamma_p t}.
\end{align}

To obtain the dynamics of the generated light only, the coherent state of the pump is traced-out from the Lindblad master equation in Eq. \eqref{eq:lindblad}. This amounts to replacing the total density operator with the reduced density operator $\bra{\alpha(t)} \rho(t) \ket{\alpha(t)}$ and replacing the nonlinear Hamiltonian in Eq. \eqref{eq:HNL} with its form in the undepleted pump approximation \cite{quantumopticsGarrison}, given by 
\begin{align}
    \label{eq:Hundepeleted}
   H_{\rm NL}&=\sum_{m,l,p_1,p_2 }G_{mlp_1p_2}\alpha_{p_1} \alpha_{p_2} {\rm e}^{-i(\tilde{\omega}_{p_1} + \tilde{\omega}_{p_2})t}b^\dagger_{m} b^\dagger_{l}  + {\rm H.c.},
\end{align} 
where the pump operators $b_{p_1}$ and $b_{p_2}$ were replaced with their the expectation value using the coherent state. We now make the crucial assumption that the pump is in a single mode $P$. Putting $p_1 = p_2 \equiv P$ into Eq. \eqref{eq:Hundepeleted}, we obtain
 \begin{align}
    \label{eq:Hundepeleted1}
    H_{\rm NL} &=\alpha^2(t) \sum_{m,l }G_{mlPP}b^\dagger_{m} b^\dagger_{l} + {\rm H.c.},
\end{align}
where 
\begin{align}
    \label{eq:pumpfieldtime}
   \alpha(t)&= \alpha_P{\rm e}^{-i \tilde{\omega}_Pt},
\end{align}
is the time-dependent pump amplitude. For example in the CROW in Fig. \ref{fig:CROW}, this assumption corresponds to the pump being in a single Bloch mode. For some systems, however, this assumption can be relaxed. If, for example, one were to take a Gaussian pump pulse that was normal to the surface of the CROW (as was done in Ref. \cite{Seifoory2019counterpropagatingCV}), then the pump would be in a continuum of free-space modes.  However, if the transverse profile of the pump inside the crystal slab does not depend on frequency, then only a single pump mode is important to the nonlinear interaction. Another example is where one considers a pulsed pump in the continuum of modes of the channel waveguide in Fig. \ref{fig:ring}. If the duration of the pump pulse in the channel is much longer than the ring round-trip time, then the pump fits mostly into a single mode of the ring (as we have shown in Ref. \cite{onemodeSL}). Since we neglect the nonlinear interaction in the channel and only consider generation in the ring, then only a single pump mode is crucial to the nonlinear interaction in the ring. In both of the last two examples we can replace the constant pump amplitude $\alpha_P$ in Eq. \eqref{eq:pumpfieldtime} with a slowly-varying temporal envelope $\alpha_P(t)$. The crucial point is that the nonlinear Hamiltonian must take the form in Eq. \eqref{eq:Hundepeleted1} in order to derive our solution to the Lindblad master equation. 

Note that, although we are using the undepleted pump approximation, this formalism does include linear pump loss, through the imaginary parts of the pump mode frequencies. Note also that one can easily adapt this formalism for SPDC, and all the results below will follow, by just replacing $\alpha(t)^2$ with a single pump amplitude $\alpha(t)$ and using the nonlinear parameter $\chi^{(2)}$   for a second-order nonlinear process \cite{Seifoory2019counterpropagatingCV} in Eq. \eqref{eq:Hundepeleted1} instead.

In the next section we derive a semi-analytic solution to the Lindblad master equation for the Hamiltonian given by Eqs. \eqref{eq:systemH} and \eqref{eq:Hundepeleted1} .
   
\section{Analytic solution to the Lindblad master equation}
\label{solution}
In previous work we studied squeezed light generation in two-mode lossy cavities. We showed that the density operator for the generated light in the cavity is a two-mode squeezed thermal state for all time \cite{twomodeSL}.  In this work we show that for a structure with many lossy quasimodes, the density operator for the generated light is the \textit{multimode squeezed thermal state} (MSTS),
\begin{equation}
    \label{eq:MSTS}
    \rho(t) = S\left(t\right)\rho_{\rm th}(t)S^\dagger\left(t\right).
\end{equation}
Here, $S(t)$ is the unitary multimode squeezing operator given by \cite{multimodeSqueezeOp}
\begin{align}
    \label{eq:squeezeoperator}
    S(t) = \exp\left(\frac{1}{2}\sum_{m,l }z_{ml}(t)b^\dagger_{m}b^\dagger_{l}-{\rm H.c.}\right),
\end{align}
where $z_{ml}(t)$ are the elements of the complex symmetric squeezing matrix $z(t)$. The multimode thermal state $\rho_{\rm th}(t)$ is a product of single-mode thermal states in each mode \cite{squeezedthermal}, such that
\begin{align}
    \label{eq:multithermalstate}
    \rho_{\rm th}(t) &= \prod_m \frac{1}{1+n_m(t)} \left(\frac{n_m(t)}{1+n_m(t)}\right)^{b^\dagger_m b_m},
\end{align}
where $n_m(t)$ is defined as the average thermal photon number for the $m$th mode. We stress that the thermal photons are not related to actual thermal effects, but rather capture the process of photon loss. The presence of $n_m(t)$ is due to a scattering process that breaks the entanglement between the generated signal and idler photon pairs. At this point the matrix $z(t)$ and functions $n_m(t)$ are unknown functions of time; in Sec. \ref{takagi} and Sec. \ref{differentialequations} we will derive equations of motion for them. 

 To show that the MSTS in Eq. \eqref{eq:MSTS} is the solution to the Lindblad master equation in Eq. \eqref{eq:lindblad}, we require that the equality
\begin{align}
\label{eq:solution0}
  I&=\rho_{\rm th}^{-1/2}(t)S^\dagger (t)\rho(t) S(t)\rho_{\rm th}^{-1/2}(t)
\end{align}
is satisfied for all time, where $I$ is the identity operator. Taking the time derivative of Eq. \eqref{eq:solution0}, it can be shown that
\begin{align}
\label{eq:solution1}
  0&=-\rho_{\rm th}^{-1/2}S^\dagger\frac{dS}{dt}\rho_{\rm th}^{1/2}+\rho_{\rm th}^{1/2}S^\dagger\frac{dS}{dt}\rho_{\rm th}^{-1/2} \nonumber
  \\
  &+\rho_{\rm th}^{-1/2}S^\dagger\frac{d\rho}{dt}S\rho_{\rm th}^{-1/2}+2\frac{d\rho_{\rm th}^{-1/2}}{dt}\rho_{\rm th}^{1/2},
\end{align}
where we drop the time-dependence for convenience. 

The majority of the rest of this section is devoted to simplifying the four terms on the right-hand side of Eq. \eqref{eq:solution1}.   

Let the  argument of the exponential squeezing operator be 
\begin{align}
    \label{eq:sigma_qms}
    \sigma \equiv \frac{1}{2}\sum_{m,l}z_{ml}b^\dagger_{m}b^\dagger_{l}-{\rm H.c.},
\end{align}
such that $S = \exp(\sigma)$ and $S^\dagger = \exp(-\sigma)$. In order to simplify Eq. \eqref{eq:solution1} we need to know  the time derivative of the squeezing operator of Eq. \eqref{eq:squeezeoperator}. This is not straightforward since $\sigma$ does not commute with its time derivative. The derivative of the squeezing operator can be written as
\begin{align}
    \label{eq:derivativeofS00}
     \frac{dS}{dt}&=\frac{d}{dt}\left(1 + \sigma + \frac{\sigma^2}{2!}+\frac{\sigma^3}{3!}+\ldots\right) \nonumber
     \\
     &=\sum_{n=0}^\infty \sum_{k=0}^\infty \frac{\sigma^n\dot{\sigma}\sigma^k}{(n+k+1)!},
\end{align}
where $\dot{\sigma}\equiv d\sigma/dt$. The  sum in the last line of Eq. \eqref{eq:derivativeofS00} has the integral representation
\begin{align}
    \label{eq:integralrep}
     \sum_{n=0}^\infty \sum_{k=0}^\infty \frac{\sigma^n\dot{\sigma}\sigma^k}{(n+k+1)!}&=\int_0^1dy {\rm e}^{(1-y) \sigma}\dot{\sigma}{\rm e}^{y \sigma}.
 \end{align}
To prove this, we expand the exponentials in the integral in a series
\begin{align}
    \label{eq:proof0}
    \int_0^1dy {\rm e}^{(1-y) \sigma}\dot{\sigma}{\rm e}^{y \sigma}&=\sum_{n=0}^\infty \sum_{k=0}^\infty\frac{\sigma^n \Dot{\sigma} \sigma^k}{n!k!}\int_0^1 dy  (1-y)^ny^k \nonumber
    \\
   &=\sum_{n=0}^\infty \sum_{k=0}^\infty \frac{\sigma^n\dot{\sigma}\sigma^k}{(n+k+1)!},
\end{align}
where the last line follows from the Euler integral of the first kind,
\begin{align}
    \label{eq:proofint}
 \int_0^1 dy  (1-y)^n y^k&=\frac{n!k!}{(n+k+1)!}.
\end{align}
Thus, putting Eq. \eqref{eq:integralrep} into Eq. \eqref{eq:derivativeofS00} we obtain
\begin{align}
    \label{eq:derivativeofS0}
    \frac{dS}{dt}&= \int_0^1dy {\rm e}^{(1-y) \sigma}\dot{\sigma}{\rm e}^{y \sigma}.
\end{align}
Using the well-known Baker-Campbell-Hausdorff formula \cite{BCH} on the integrand in Eq. \eqref{eq:derivativeofS0}  and multiplying the equation by $S^\dagger=\exp(-\sigma)$ from the left, we obtain
\begin{align}
    \label{eq:derivativeofS}
    S^\dagger\frac{dS}{dt}&=\sum_{k=0}^{\infty}\frac{(-1)^k}{(k+1)!}\mathcal{S}_k,
\end{align}
where the terms $\mathcal{S}_k$ can be obtained recursively from
\begin{align}
    \label{derivativeofS_terms}
    \mathcal{S}_{k+1} = [\sigma, \mathcal{S}_{k}],
\end{align}
for $k=0,1,\ldots$, where
\begin{align}
    \label{eq:derivativeofS_term0}
    \mathcal{S}_0 &\equiv  \dot{\sigma}.
\end{align}
Each  $\sigma$ or  $\dot{\sigma}$ that appears in Eqs. \eqref{derivativeofS_terms} or \eqref{eq:derivativeofS_term0}  involves a double sum over the quasimodes (see Eq. \eqref{eq:sigma_qms}). Thus, calculating the $k>0$ terms of $\mathcal{S}_k$ becomes intractable for multiple modes since each term has $2(k+1)$ sums.
However, by introducing a new Schmidt basis via a Takagi factorization of $H_{\rm NL}$, in which the squeezing operator is diagonal, we are able to calculate the terms $\mathcal{S}_k$ and show that the sum in Eq. \eqref{eq:derivativeofS} converges even for the case of a multimode squeezing operator (see Eq. \eqref{eq:derivativeofSdiagonalfinal}).  In the next subsection we introduce this new Schmidt basis. 
\subsection{The Schmidt basis and the diagonal multimode squeezing operator}
\label{takagi}
In this subsection we diagonalize the nonlinear Hamiltonian in Eq. \eqref{eq:Hundepeleted1} and the multimode squeezing operator in Eq. \eqref{eq:squeezeoperator} so that we can simplify the four terms in Eq. \eqref{eq:solution1} and prove that the MSTS is the solution to the Lindblad master equation. 

We start by performing a Takagi factorization of the nonlinear parameter $G$ in $H_{\rm NL}$. The Takagi factorization decomposes the complex symmetric square matrix $G$ into the form
 \begin{align}
     \label{eq:takagi}
     G= U\Lambda U^{\rm T},
 \end{align}
 where $\Lambda = {\rm diag}(\lambda_1,\lambda_2,\ldots)$ is a diagonal matrix with complex entries, $U$ is a unitary matrix with $U^\dagger U =1$ and $U^{\rm T}$ is the transpose of $U$. The Takagi factorization is a special case of symmetric singular value decomposition (SVD) \cite{divideConquerTakagi}. In Appendix \ref{SVD+takagi} we show that the diagonal matrix $\Lambda$ from the Takagi factorization is just a scaled version of the matrix of singular values  obtained from the symmetric SVD. 
 
Putting Eq. \eqref{eq:takagi} into  Eq. \eqref{eq:Hundepeleted1} we obtain the diagonalized nonlinear Hamiltonian
\begin{equation}
    \label{eq:HNLdiagonal}
    H_{\rm NL} = \alpha^2(t)\sum_\mu \lambda_\mu B^{\dagger 2}_\mu + {\rm H.c.},
\end{equation}
where $\lambda_\mu$ ($\mu = 1,2,\ldots$) are the diagonal entries of $\Lambda$, and we define the new creation and annihilation Schmidt operators
\begin{align}
    \label{eq:Schmidtoperator}
    B^\dagger_\mu &\equiv \sum_m U_{ m \mu}b^\dagger_m,
    \\
    B_\mu &\equiv \sum_m U^*_{ \mu m}b _m,
\end{align}
that have the standard commutation relation
\begin{align}
    \label{eq:Schmidtcommutator}
    \left[B_\mu,B^\dagger_\nu\right] = \delta_{\mu\nu}
\end{align}
due to the orthogonality of the basis $(\sum_{m} U^*_{\mu m}U_{m\nu} = \delta_{\mu \nu})$. We call the $B_\mu$  Schmidt operators, since when there is no loss they give the Schmidt decomposition of the multimode squeezed state. The inverse transformation is
\begin{align}
    \label{eq:reversetransformation}
    b_m&=\sum_\mu U_{m \mu} B_\mu.
\end{align}

It is clear now why we have to assume the pump is separable into a spatial and temporal part (see $H_{\rm NL}$ in Eq. \eqref{eq:Hundepeleted1}). This makes the Schmidt mode basis independent of time. Otherwise, a new Takagi factorization would have to be done at each time and the operators $B_\mu$ would become time-dependent. In practice one calculates the nonlinear parameters $G$ for a given structure by first calculating the quasimodes of the structure using a method such as finite-difference time-domain (FDTD) and then calculating the spatial overlap given in Eq. \eqref{eq:nonlinearparameter}. Then one performs the Takagi factorization of  $G$ to obtain the matrices $U$ and $\Lambda$. 

We now make the ansatz for the squeezing parameter matrix $z(t)$, 
\begin{align}
    \label{eq:takagisqueezingoparam}
    z(t) = U r(t) {\rm e}^{i\phi(t)} U^{\rm T},
\end{align}
where $r(t)$ and $\phi(t)$ are real diagonal matrices containing the squeezing amplitudes and  squeezing phases of the Schmidt modes, that is $r(t) = \text{diag}(r_1(t), r_2(t),\ldots)$ and $\phi(t) = \text{diag}(\phi_1(t), \phi_2(t),\ldots)$. Putting Eq. \eqref{eq:takagisqueezingoparam} into Eq. \eqref{eq:squeezeoperator}, we obtain the diagonal multimode squeezing operator
\begin{align}
    \label{eq:diagonalsqueezeoperator}
    S(t) &= \prod_\mu S_\mu(t),
\end{align}
where $S_\mu(t)$ is the single mode squeezing operator for the $\mu$th Schmidt mode, given by
\begin{align}
    \label{eq:diagonalsqueezeoperator_singlemode}
    S_\mu(t) &\equiv \exp\left(\frac{1}{2}r_\mu(t){\rm e}^{i\phi_\mu(t)}B^{\dagger2}_{\mu}-{\rm H.c.}\right).
\end{align}
We show in Sec. \ref{differentialequations} that this is the correct form of the multimode squeezing operator, because when we impose this form, we can derive equations for the $r_\mu(t)$ and $\phi_\mu(t)$ (see Eqs. \eqref{eq:rdiff} and \eqref{eq:phidiff}) that describe a MSTS that is a solution to the Lindblad master equation. Note, one can obtain the squeezing matrix in the quasimode basis $z(t)$, after solving the equations for the $r_\mu(t)$ and $\phi_\mu(t)$ and putting them into Eq. \eqref{eq:takagisqueezingoparam}.

Before moving to the next subsection, where we take the time derivative of $S(t)$, we present the squeezing transformation for the $B_\mu$ and $b_m$ operators. These transformations are necessary when squeezing the terms in Eq. \eqref{eq:solution1}. Using the squeezing operator in Eq. \eqref{eq:diagonalsqueezeoperator}, we obtain
\begin{align}
    \label{eq:Btransform}
    S^\dagger B_\mu S&= S_\mu^\dagger B_\mu S_\mu \nonumber
    \\
    &=\cosh(r_\mu)B_\mu - {\rm e}^{i\phi_\mu}\sinh(r_\mu)B^\dagger_\mu,
\end{align}
where the time-dependence is dropped for convenience.
All other squeezing transformations involving products of Schmidt operators (such as $B_\mu B_\nu$) can be derived from Eq. \eqref{eq:Btransform} and the fact that different Schmidt operators commute. The squeezing transformation of the $b_m$ operators is derived with the help of the inverse transformation in Eq. \eqref{eq:reversetransformation}. For the operator $b_m$, we obtain
\begin{align}
\label{eq:btransform}
    S^\dagger b_mS&=\sum_\mu U_{m \mu}S_\mu^\dagger B_\mu S_\mu.
\end{align}

\subsection{The time derivative of the multimode squeezing operator}
\label{squeezeopderivative}
In this subsection we simplify the first two terms in Eq. \eqref{eq:solution1} that involve the derivative of the squeezing operator. We show that using the diagonal squeezing operator in Eq. \eqref{eq:diagonalsqueezeoperator}, we can calculate the sum in Eq. \eqref{eq:derivativeofS}. 

Using Eq. \eqref{eq:diagonalsqueezeoperator}, along with the fact that the $S_\mu$ commute and that $S^\dagger_\mu S_\mu = I$, the left-hand side of Eq. \eqref{eq:derivativeofS} can be written as
\begin{align}
    \label{eq:derivativeofSdiagonal00}
   S^\dagger\frac{dS}{dt}&=\sum_\mu S_\mu^\dagger\frac{dS_\mu}{dt}.
\end{align}
Each term in the sum in Eq. \eqref{eq:derivativeofSdiagonal00} can be expanded using the right-hand side of Eq. \eqref{eq:derivativeofS}
\begin{align}
    \label{eq:derivativeofSdiagonal0}
 S_\mu^\dagger\frac{dS_\mu}{dt}&=\sum_{k=0}^{\infty}\frac{(-1)^k}{(k+1)!}\mathcal{S}_{k\mu}.
\end{align}
Putting Eq. \eqref{eq:derivativeofSdiagonal0} into Eq. \eqref{eq:derivativeofSdiagonal00}, we obtain
\begin{align}
    \label{eq:derivativeofSdiagonal}
   S^\dagger \frac{dS}{dt}
   &=\sum_\mu\sum_{k=0}^{\infty}\frac{(-1)^k}{(k+1)!}\mathcal{S}_{k\mu},
\end{align}
where the terms $\mathcal{S}_{k\mu}$ in Eq. \eqref{eq:derivativeofSdiagonal} can be obtained recursively from 
\begin{align}
\label{eq:k>0terms}
\mathcal{S}_{k+1\mu} = [\sigma_\mu,\mathcal{S}_{k\mu}]
\end{align}
for $k=0,1,\ldots$, where the $k=0$ term is defined as
\begin{align}
    \label{eq:derivativeofS_diagonal1}
    \mathcal{S}_{0\mu} &\equiv  \dot{\sigma}_\mu, 
\end{align}
where
\begin{align}
\label{eq:sigma_mu}
\sigma_\mu \equiv \frac{1}{2} r_\mu \exp(i\phi_\mu)B^{\dagger 2}_\mu - {\rm H.c.}.
\end{align}
 Since the definition of $\sigma_\mu$ does not contain sums over Schmidt modes, the problem is reduced to just evaluating a series of commutators involving only single mode operators. This is straightforward to do using the commutation relation for the $B_\mu$ operators. The odd $k$ and even $k$ terms of the sum are
\begin{align}
\label{eq:termsofsum}
     \mathcal{S}_{k\mu}& = \begin{cases} -\frac{i}{4}(2r_\mu)^{k+1}\dot{\phi}_\mu\left(2B^\dagger_\mu B_\mu+1 \right), \,\,\,\,\, {\rm odd}\, k\ge 1
     \\
     \\
 \frac{i}{4}(2r_\mu)^{k+1}\dot{\phi}_\mu\left({\rm e}^{i\phi_\mu}B^{\dagger2}_\mu +{\rm H.c.} \right), \,\,\,\,\, {\rm even}\, k\ge 2.
    \end{cases}
\end{align}
Putting $\Dot{\sigma}_\mu$ and the terms in Eq. \eqref{eq:termsofsum} into Eq. \eqref{eq:derivativeofSdiagonal} and summing over $k$, we obtain
\begin{align}
    \label{eq:derivativeofSdiagonalfinal}
   S^\dagger \frac{dS}{dt}&=\sum_\mu\Bigg( \frac{i}{2}\sinh^2(r_\mu)\frac{d\phi_\mu}{dt}\left(2B^\dagger_\mu B_\mu+1 \right) \nonumber
    \\
    &+\left(\frac{1}{2}\frac{d r_\mu}{dt}+\frac{i}{4}\sinh(2r_\mu)\frac{d\phi_\mu}{dt} \right){\rm e}^{i\phi_\mu}B^{\dagger 2}_\mu \nonumber
    \\
    &-\left(\frac{1}{2}\frac{d r_\mu}{dt}-\frac{i}{4}\sinh(2r_\mu)\frac{d\phi_\mu}{dt} \right){\rm e}^{-i\phi_\mu}B^{ 2}_\mu \Bigg).
\end{align}
In order to derive Eq. \eqref{eq:derivativeofSdiagonalfinal} it is necessary that the Schmidt operators are time-independent and have the standard commutator. In the next subsection we put Eq. \eqref{eq:derivativeofSdiagonalfinal} into Eq. \eqref{eq:solution1} and derive the set of coupled first-order differential equations for the squeezing amplitudes ($r_\mu$), squeezing phases ($\phi_\mu$), and thermal photon numbers ($n_m$).
\subsection{Differential equations for the squeezing parameters and thermal photon numbers}
\label{differentialequations}
We are now in a position to simplify the four terms in Eq. \eqref{eq:solution1}. To make the derivation clearer we redefine the terms in Eq. \eqref{eq:solution1} to obtain 
\begin{align}
\label{eq:maineq}
0=T1+T2+T3,    
\end{align}
where 
\begin{align}
    \label{eq:term1}
    T1&\equiv-\rho_{\rm th}^{-1/2}S^\dagger(z)\frac{dS(z)}{dt}\rho_{\rm th}^{1/2}+\rho_{\rm th}^{1/2}S^\dagger(z)\frac{dS(z)}{dt}\rho_{\rm th}^{-1/2}, 
\\
    \label{eq:term2}
    T2&\equiv\rho_{\rm th}^{-1/2}S^\dagger(z)\frac{d\rho}{dt}S(z)\rho_{\rm th}^{-1/2},
\\
    \label{eq:term3}
    T3&\equiv 2\frac{d\rho_{\rm th}^{-1/2}}{dt}\rho_{\rm th}^{1/2}.
\end{align}
The transformations involving $\rho_{\rm th}^{\pm 1/2}$ in $T1$ and $T2$ are performed using the identity
\begin{align}
    \label{eq:thermaltransform}
    \rho_{\rm th}^{\pm 1/2} B_\mu \rho_{\rm th}^{\mp 1/2}&=\sum_m U^*_{m\mu}b_m x_m^{\mp1/2},
\end{align}
where
\begin{align}
    \label{eq:xtrans}
    x_m \equiv \frac{n_m}{1+n_m}.
\end{align}
It is simple to generalize this identity to transformations involving the product of operators, such as $\rho_{\rm th}^{\pm 1/2} B_\mu B_\nu \rho_{\rm th}^{\mp 1/2}$. 

Our strategy is to reduce $T1$, $T2$, and $T3$ to expressions that are a sum of Schr\"{o}dinger operators multiplied by time-dependent coefficients that depend on the squeezing amplitudes $r_\mu(t)$, squeezing phases $\phi_\mu(t)$, and thermal photon numbers $n_m(t)$, respectively as well as their first derivatives. Then we force these coefficients to be zero, such that Eq. \eqref{eq:maineq} is satisfied. The result is a set of coupled first-order differential equations for $\dot{r}_\mu(t)$, $\dot{\phi}_\mu(t)$, and $\dot{n}_m(t)$.

Putting Eq. \eqref{eq:derivativeofSdiagonalfinal} in $T1$ and using Eq. \eqref{eq:thermaltransform}, we obtain
\begin{align}
    \label{eq:term1_2}
    T1&=\sum_{m,l}\left(D_{ml}b^\dagger_mb^\dagger_l + D^*_{ml}b_m b_l + F_{ml}b^\dagger_m b_l\right),
\end{align}
where the time-dependent coefficients $D_{ml}$ and $F_{ml}$ are
\begin{align}
    \label{eq:Dcoeff}
    D_{ml}&= \frac{x_mx_l - 1}{2\sqrt{x_mx_l}}\sum_\mu U_{m\mu}U_{l\mu} \left(\dot{r}_\mu+\frac{i}{2}\sinh(2r_\mu)\dot{\phi}_\mu \right){\rm e}^{i\phi_\mu} ,
\end{align}
and
\begin{align}
    \label{eq:Ncoeff}
    F_{ml}&=   i\,\frac{x_m - x_l}{\sqrt{x_mx_l}}\sum_\mu U_{m\mu}U^*_{l\mu} \dot{\phi}_\mu \sinh^2(r_\mu) .
\end{align}
Obtaining these coefficients requires many lines of algebra but it is a straightforward exercise.

In order to derive the equations for $\dot{r}_\mu$ and $\dot{\phi}_\mu$, we define new Schr\"{o}dinger operators 
\begin{align}
    \label{eq:Vop}
    \hat{V}_{ml}&\equiv b_mb_l + b^\dagger_m b^\dagger_l, 
    \\
     \label{eq:Wop}
    \hat{W}_{ml}&\equiv -ib_mb_l + ib^\dagger_m b^\dagger_l, 
\end{align}
such that  Eq. \eqref{eq:term1_2} can be written as
\begin{align}
    \label{eq:term1_3}
        T1&=\sum_{m,l}\left(\Re{D_{ml}}\hat{V}_{ml} + \Im{D_{ml}}\hat{W}_{ml} + F_{ml}b^\dagger_m b_l\right),
\end{align}
where $\Re{D_{ml}}$ and $\Im{D_{ml}}$ are the real and imaginary parts of $D_{ml}$. 

Next we write $T2$ in terms of the operators $\hat{V}_{ml}$ and $\hat{W}_{ml}$. The main calculation in $T2$ is the squeezing transformation of $d\rho/dt$. This is the Lindblad master equation in Eq. \eqref{eq:lindblad}. The nonlinear Hamiltonian in the master equation is replaced with its diagonal form in Eq. \eqref{eq:HNLdiagonal}. The squeezing transformations are performed using Eq. \eqref{eq:btransform}. Then by using Eq. \eqref{eq:thermaltransform}, it can be shown that Eq. \eqref{eq:term2} simplifies to
\begin{align}
    \label{eq:term2_2}
    T2&=\sum_{m,l}\left(\Re{E_{ml}}\hat{V}_{ml} + \Im{E_{ml}}\hat{W}_{ml}+K_{ml}b^\dagger_mb_l\right) \nonumber
    \\
     &+ \sum_{\mu, \nu}\Gamma_{\mu \nu} \cosh(r_\mu)\cosh(r_\nu)\sum_m U_{m\mu}U^*_{m\nu}x_m\nonumber
     \\
     &-\sum_\mu \Gamma_{\mu \mu} \sinh^2(r_\mu),
\end{align}
where
\begin{align}
    \label{eq:transformed_decayrates}
\Gamma_{\mu \nu}\equiv 2\sum_m \gamma_m U^*_{m \mu}U_{m \nu},
\end{align}
and the time-dependent coefficients $E_{ml}$ and $K_{ml}$ are
\begin{widetext}
\begin{align}
\label{eq:Ecoeff}
    E_{ml}&=\sum_{\mu,\nu}U_{m\mu}U_{l\nu}{\rm e}^{i\phi_\nu}\cosh(r_\mu)\sinh(r_\nu)\left(-i\Omega_{\mu \nu} \frac{x_mx_l-1}{\sqrt{x_mx_l}}+\frac{1}{2}\Gamma_{\mu \nu} \frac{1+x_mx_l-2x_m}{\sqrt{x_mx_l}} \right) \nonumber
    \\
    &+ \frac{i}{\hbar}\,\frac{x_mx_l-1}{\sqrt{x_mx_l}}\sum_\mu U_{m\mu}U_{l\mu}{\rm e}^{i\phi_\mu} \left(\alpha^2\lambda_\mu{\rm e}^{-i\phi_\mu} \cosh^2(r_\mu) + \alpha^{*2}\lambda^*_\mu {\rm e}^{i\phi_\mu}\sinh^2(r_\mu)\right), 
    \\
    \label{eq:Gcoeff}
    K_{ml}&=\sum_{\mu,\nu}U_{m\mu}U^*_{l\nu}\cosh(r_\mu)\cosh(r_\nu)\left(i\Omega_{\mu \nu}\frac{x_m-x_l}{\sqrt{x_mx_l}}+\frac{1}{2}\Gamma_{\mu \nu}\frac{2x_mx_l-x_m-x_l}{\sqrt{x_mx_l}}\right) \nonumber
    \\
    &+\sum_{\mu,\nu}U_{m\mu}U^*_{l\nu}{\rm e}^{i(\phi_\mu - \phi_\nu)}\sinh(r_\mu)\sinh(r_\nu)\left(i\Omega^*_{\mu \nu}\frac{x_m-x_l}{\sqrt{x_mx_l}}+\frac{1}{2}\Gamma^*_{\mu \nu}\frac{2-x_m-x_l}{\sqrt{x_mx_l}}\right) \nonumber
    \\
    &-\frac{2}{\hbar}\frac{x_m-x_l}{\sqrt{x_mx_l}}\sum_\mu U_{m\mu}U^*_{l\mu}\cosh(r_\mu)\sinh(r_\mu) \left(\alpha^2\lambda_\mu{\rm e}^{-i\phi_\mu} + \alpha^{*2}\lambda^*_\mu {\rm e}^{i\phi_\mu}\right),
\end{align}
\end{widetext}
where 
\begin{align}
    \label{transformed_freq}
    \Omega_{\mu \nu}\equiv \sum_m \omega_m U^*_{m \mu}U_{m \nu}.
\end{align} 

Since the derivative of the thermal state $\rho_{\rm th}$ in Eq. \eqref{eq:multithermalstate} is easy, $T3$ requires no special treatment in order to simplify. Putting Eq. \eqref{eq:multithermalstate} into Eq. \eqref{eq:term3} we obtain
\begin{align}
    \label{eq:term3_2}
    T3&=\sum_m\left(-\frac{\dot{x}_m}{x_m}b^\dagger_mb_m +\frac{\dot{x}_m}{1-x_m}\right).
\end{align}

To form the differential equations for $\dot{r}_\mu$ and $\dot{\phi}_\mu$, we set the sum of the coefficients in front of $\hat{V}_{ml}$ and $\hat{W}_{ml}$ in Eqs. \eqref{eq:term1_3} and \eqref{eq:term2_2}  to zero. This gives
\begin{align}
    \label{eq:DandE_real}
  \Re{D_{ml}}&= - \Re{E_{ml}},
  \\
   \label{eq:DandE_imag}
  \Im{D_{ml}}&= - \Im{E_{ml}}.
\end{align}
In Appendix \ref{appndx1}, we show that Eq. \eqref{eq:DandE_real} leads to an equation for $\dot{r}_\mu$ and Eq. \eqref{eq:DandE_imag} for $\Dot{\phi}_\mu$. The resulting differential equations are
\begin{widetext}
\begin{align}
    \label{eq:rdiff}
    \Dot{r}_\mu&=\frac{1}{i\hbar}\left(\alpha^2\lambda_\mu{\rm e}^{-i\phi_\mu} - {\rm c.c.}\right)-\frac{1}{2}\sum_{\nu,\sigma}\cosh(r_\nu)\sinh(r_\sigma)\left(\Gamma_{\nu \sigma}{\rm e}^{i(\phi_\sigma - \phi_\mu)}\sum_{m,l}\frac{-n_m+n_l+1}{n_m+n_l+1}U_{m\nu}U_{l\sigma}U^*_{m\mu}U^*_{l\mu}+{\rm c.c.}\right),
\\
    \label{eq:phidiff}
    \Dot{\phi}_\mu&=2\Omega_{\mu\mu}-\frac{2}{\hbar\tanh(2r_\mu)}\left(\alpha^2\lambda_\mu{\rm e}^{-i\phi_\mu} + {\rm c.c.}\right)\nonumber
    \\
    &+\frac{i}{2}\sum_{\nu,\sigma}\frac{\cosh(r_\nu)\sinh(r_\sigma)}{\cosh(r_\mu)\sinh(r_\mu)}\left(\Gamma_{\nu \sigma}{\rm e}^{i(\phi_\sigma - \phi_\mu)}\sum_{m,l}\frac{-n_m+n_l+1}{n_m+n_l+1}U_{m\nu}U_{l\sigma}U^*_{m\mu}U^*_{l\mu}-{\rm c.c.}\right).
\end{align}
\end{widetext}

In order to obtain an equation for $\Dot{n}_m$, the operators $b^\dagger_m b_l$ in $T1$, $T2$, and $b^\dagger_m b_m$ in $T3$ are expanded in terms of the Schmidt operator $B^\dagger_\mu B_\nu$ using Eq. \eqref{eq:reversetransformation}.
This is necessary since it creates a common operator $B^\dagger_\mu B_\nu$ that is shared between $T1$, $T2$, and $T3$. Then setting the sum of the coefficients in front $B^\dagger_\mu B_\nu$ equal to zero, we show in Appendix \ref{appndx2} that this leads to the following equation for $\Dot{n}_m$, 
\begin{widetext}
\begin{align}
    \label{eq:ndiff}
    \Dot{n}_m&=(1+n_m )\sum_{\nu,\sigma}U_{m\nu}U^*_{m\sigma}\Gamma_{\nu \sigma}^*{\rm e}^{i(\phi_\nu -\phi_\sigma)} \sinh(r_\nu)\sinh(r_\sigma)- n_m\sum_{\nu,\sigma}U_{m\nu}U^*_{m\sigma}\Gamma_{\nu \sigma}\cosh(r_\nu)\cosh(r_\sigma).
\end{align}
\end{widetext}
In order to complete the derivation, we have to show that the sum of the coefficients in front of the identity operator in $T2$ (Eq. \eqref{eq:term2_2}) and $T3$ (Eq. \eqref{eq:term3_2}) is zero. Collecting the appropriate terms from Eq. \eqref{eq:term2_2} and Eq. \eqref{eq:term3_2}, we obtain
\begin{align}
    \label{eq:identity_sum_equation}
    0&=\sum_m \frac{\Dot{n}_m}{1+n_m} -\sum_\mu \Gamma_{\mu \mu} \sinh^2(r_\mu) \nonumber
     \\
     &+\sum_{\mu, \nu}\Gamma_{\mu \nu} \cosh(r_\mu)\cosh(r_\nu)\sum_m U_{m\mu}U^*_{m\nu}\frac{n_m}{1+n_m}.
\end{align}
Putting  Eq. \eqref{eq:ndiff} into Eq. \eqref{eq:identity_sum_equation} and summing over $m$, it is easily shown that the right-hand side is equal to zero, where we use the fact that $\sum_m U_{m\nu}U^*_{m\sigma} = \delta_{\nu \sigma}$, and $\Gamma^*_{\mu \mu} = \Gamma_{\mu \mu}$. This completes the derivation since we have shown that the coefficients in front of the operators $\hat{V}_{ml}$, $\hat{W}_{ml}$, $B^\dagger_\mu B_\nu$, and the identity operator sum to zero.

The key result of this work are Eqs. \eqref{eq:rdiff} - \eqref{eq:ndiff}, which describe the time-dependence of the squeezing amplitudes, squeezing phases, and thermal photon numbers of the MSTS. For a system of $M$ quasimodes, they form a set of $3M$ coupled first-order differential equations, which can be easily solved on a standard PC. The benefit of having these equations is that once they are solved, we have the time-dependent density operator of the MSTS and with it we can calculate any observables such as the correlation variance and photon number. Note that the number of equations does not depend on the number of photons at all. In contrast, numerical techniques for calculating the density operator often require solving a large number of coupled equations that depend on the photon number \cite{numericalLindblad}. Therefore our approach can greatly reduce the number of coupled equations, making it more feasible to study large, multimode, lossy structures.

We note that the coupled equations as formulated apply only to a set of discrete modes. If one has a continuum of modes, such as for the Bloch modes of a CROW, one needs to simply discritize the modes by applying the appropriate (\textit{e.g.} periodic) boundary conditions. Otherwise, the equations become a set of coupled integro-differential equations, which become difficult to solve computationally without discretization. In Sec. \ref{results} we solve these equations for a CROW with four cavities, where we use periodic boundary conditions to quantize the allowed Bloch vectors. In Appendix \ref{solving} we discuss how to solve Eqs. \eqref{eq:rdiff} - \eqref{eq:ndiff} when the initial state is the vacuum state. 

In the next section we derive expressions for the expectation value of the photon number operator and the same-time correlation variance between different quasimodes using the MSTS. 
\section{Expectation values and the correlation variance}
\label{expectation}
In this section we use the MSTS to evaluate the expectation value of the quasimode number operator and to calculate the same-time correlation variance between pairs of quasimodes. Let $X_m(\varphi_m)$ and $Y_m(\varphi_m)$ be orthogonal quadrature operators defined as
\begin{align}
    \label{eq:Xquad}
    X_m(\varphi_m)&=\frac{b_m{\rm e}^{-i\varphi_m}+b^\dagger_m{\rm e}^{i\varphi_m}}{2},
    \\
     \label{eq:Yquad}
    Y_m(\varphi_m)&=\frac{b_m{\rm e}^{-i\varphi_m}-b^\dagger_m{\rm e}^{i\varphi_m}}{2i},
\end{align}
where $\varphi_m$ is an angle in phase space. In order to quantify the inseparability of the MSTS, the correlation variance \cite{onChipCV} between two quasimodes $m$ and $l$ is defined as 
\begin{align}
    \label{eq:corrvar}
    \Delta_{ml}^2(t)&=\left<\left[\Delta\left(X_m \pm X_{l}\right)\right]^2\right>+\left<\left[\Delta\left(Y_m \mp Y_{l}\right)\right]^2\right>,
\end{align}
where 
\begin{align}
    \label{eq:varX}
   \left< \left[\Delta\left(X_m \pm X_{l}\right)\right]^2\right> &=\left<\left(X_m \pm X_{l}\right)^2\right> \nonumber
    \\
    &\equiv {\rm Tr}\left[\rho(t)\left(X_m \pm X_{l}\right)^2 \right],
    \\
      \label{eq:varY}
   \left< \left[\Delta\left(Y_m \mp Y_{l}\right)\right]^2 \right>&=\left<\left(Y_m \mp Y_{l}\right)^2\right>\nonumber
    \\
    &\equiv {\rm Tr}\left[\rho(t)\left(Y_m \mp Y_{l}\right)^2 \right]
\end{align}
is defined as the variance of the operator and $\rho(t)$ is the density operator for the MSTS. Here we have used the fact that $\left<X_m\right> = \left<Y_m\right> =0$.
It can be proved \cite{Duan2000InseparabilitySystems,Simon2000Peres-HorodeckiSystems} that the MSTS is inseparable, and thus contains entanglement, iff $ \Delta_{ml}^2<1$. 

In order to calculate the correlation variances one needs to know the expectation values $\left<b^\dagger_m b_l\right>$ and $\left<b_m b_l\right>$. Using the Schmidt operators, we obtain
\begin{align}
\label{eq:number_ex}
\left<b^\dagger_m b_l\right>&=\sum_{\mu,\nu} U^*_{m\mu}U_{l\nu}\left<B^\dagger_\mu B_\nu\right>,
\\
\label{eq:corr_ex}
\left<b_m b_l\right>&=\sum_{\mu,\nu} U_{m\mu}U_{l\nu}\left<B_\mu B_\nu\right>,
\end{align}
where the expectation values of the Schmidt operators are 
\begin{align}
\label{eq:eigen_number_ex}
\left<B^\dagger_\mu B_\nu\right>&={\rm Tr}\left[\rho_{\rm th}(t)S^\dagger(t) B^\dagger_\mu B_\nu S(t)\right],
\\
\label{eq:eigen_corr_ex}
\left<B_\mu B_\nu\right>&={\rm Tr}\left[\rho_{\rm th}(t)S^\dagger(t) B_\mu B_\nu S(t)\right].
\end{align}
These can be simplified using the squeezing transformation in Eq. \eqref{eq:Btransform} followed by the thermal state transformation in Eq. \eqref{eq:thermaltransform}. Doing this, Eq. \eqref{eq:number_ex} and Eq. \eqref{eq:corr_ex} become
\begin{widetext}
\begin{align}
\label{eq:number_ex_final}
\left<b^\dagger_m b_l\right>&=\sum_{\mu,\nu} \left(U^*_{m\mu}U_{l\nu}\cosh(r_\mu)\cosh(r_\nu)+U^*_{m\nu}U_{l\mu}\sinh(r_\mu)\sinh(r_\nu){\rm e}^{i(\phi_\mu-\phi_\nu)}\right)\eta_{\mu \nu}+\sum_\mu U^*_{m\mu}U_{l\mu}\sinh^2(r_\mu),
\\
\label{eq:corr_ex_final}
\left<b_m b_l\right>&=-\sum_{\mu,\nu} \left(U_{m\mu}U_{l\nu}+U_{m\nu}U_{l\mu}\right)\sinh(r_\mu)\cosh(r_\nu){\rm e}^{i\phi_\mu}\eta_{\mu \nu} - \sum_\mu U_{m\mu}U_{l\mu}\cosh(r_\mu)\sinh(r_\mu){\rm e}^{i\phi_\mu},
\end{align}
\end{widetext}
where
\begin{align}
\label{eq:thmnum_eigenbasis}
    \eta_{\mu \nu}(t)&\equiv \sum_m  U_{m\mu}U^*_{m\nu}n_m(t).
\end{align}
The time-dependent expectation value of the photon number operator for the $m$th quasimode $\left<b^\dagger_mb_m\right>$ can be obtained by letting $m=l$ in Eq. \eqref{eq:number_ex_final}. Also, putting Eq. \eqref{eq:number_ex_final} and Eq. \eqref{eq:corr_ex_final} into Eq. \eqref{eq:corrvar} one can calculate the time-dependent correlation variance between any two modes and quantify the inseparability of the MSTS. 
\section{Limiting cases and comparison to other work}
\label{lmitingcases}
In this section we discuss three limiting cases to our multimode theory presented above and compare the results to other work. The first limit is when the modes are all \textit{lossless}, such that the imaginary part of every quasimode is equal to zero ($\gamma_m =0$). In this limit we show that our theory gives a  multimode squeezed vacuum state, and show that the squeezing parameter and squeezing phase we obtain agree with other work \cite{beyondPhotonPairs}. The second limit we consider is when there is only a single \textit{lossy} mode that holds the squeezed light. We show that in this limit our multimode theory gives a single-mode squeezed state in agreement  with previous work \cite{Seifoory2017SqueezedCavities}. In the third limit we allow two \textit{lossy} modes for the squeezed light and show that our theory gives a two-mode squeezed thermal state in agreement with our previous work \cite{twomodeSL}.
\subsection{Lossless modes}
\label{lossless}
If the modes are lossless, then the thermal photon number $n_m$ of each mode is equal to zero and the thermal state in Eq. \eqref{eq:MSTS} gets replaced with the vacuum state $\ket{0}\bra{0}$. Furthermore, the terms proportional to $\gamma_m$ are dropped in the Lindblad master equation, so it becomes the quantum Liouville equation. Therefore the density operator becomes
\begin{align}
    \label{eq:MSVS}
    \rho(t)&=S(t)\ket{0}\bra{0}S^\dagger(t),
\end{align}
which is the density operator for a multimode squeezed vacuum state. One can determine the squeezing parameter $r_\mu$ and squeezing phase $\phi_\mu$ for this state by letting $\gamma_m=0$ in Eq. \eqref{eq:rdiff} and Eq. \eqref{eq:phidiff}. Doing this we obtain
\begin{align}
    \label{eq:rdiff_noloss}
    \Dot{r}_\mu&=\frac{1}{i\hbar}\left(\alpha^2\lambda_\mu{\rm e}^{-i\phi_\mu} - {\rm c.c.}\right),
\\
    \label{eq:phidiff_noloss}
    \Dot{\phi}_\mu&=2\Omega_{\mu\mu}-\frac{2\left(\alpha^2\lambda_\mu{\rm e}^{-i\phi_\mu} + {\rm c.c.}\right)}{\hbar\tanh(2r_\mu)}.
\end{align}
Assuming that the system starts initially in vacuum at $t=t_i$ with $r_\mu(t_i) = 0$, the second term in Eq. \eqref{eq:phidiff_noloss} goes to infinity. To prevent this we force the numerator to be zero initially and also at all later times. Thus to obtain our final solution we put
\begin{align}
    \label{eq:initialcond_vacuum}
    \Re{\alpha^2\lambda_\mu{\rm e}^{-i\phi_\mu} }&=0.
\end{align}
Writing $\alpha^2(t) = |\alpha(t)|^2\exp(i\beta(t) )$ and $\lambda_\mu = |\lambda_\mu|\exp(i\theta_\mu)$, where $\beta(t)$ is a time-dependent phase and $\theta_\mu$ is a time-independent phase, we obtain from Eq. \eqref{eq:initialcond_vacuum}  the phase condition
\begin{align}
    \label{eq:initialcond_vacuum_phases}
    \beta(t)+\theta_\mu-\phi_\mu(t)&= \pi/2.
\end{align}
Putting Eq. \eqref{eq:initialcond_vacuum_phases} into Eq. \eqref{eq:rdiff_noloss} and Eq. \eqref{eq:phidiff_noloss} and integrating gives
\begin{align}
    \label{eq:r_noloss}
    r_\mu(t)&= \frac{2|\lambda_\mu|}{\hbar}\int_{t_i}^t dt'|\alpha(t')|^2,
\\
    \label{eq:phi_noloss}
    \phi_\mu(t)&=2(t-t_i)\Omega_{\mu\mu}+\phi_\mu(t_i).
\end{align}
Therefore the multimode squeezing operator can be written as
\begin{align}
    \label{eq:diagonalsqueezeoperator_noloss}
    S(t) &= \prod_\mu\exp\left(\frac{|\lambda_\mu|}{\hbar}r_\mu(t){\rm e}^{-2i\Omega_{\mu\mu}t}B^{\dagger2}_{\mu}-{\rm H.c.}\right),
\end{align}
where we let $\phi_\mu (t_i) = 2 \Omega_{\mu \mu}t_i$. The squeezed vacuum state we obtain is given by $S(t)\ket{0}$. This is the same state obtained by Quesada N. \textit{et al.}, following a similar procedure that uses the Takagi factorization (see Eq. (204) and Eqs. (237) - (238) in Ref. \cite{beyondPhotonPairs}). We note that this state has the same form one obtains in the weak pump limit (\textit{i.e.} $\alpha \ll 1$) by keeping only the  first-order terms in the Dyson or Magnus expansion of the evolution operator, a result that is also noted by Quesada N. \textit{et al.} \cite{beyondPhotonPairs}.   
For lossless modes the expectation values of Eq. \eqref{eq:number_ex_final} and Eq. \eqref{eq:corr_ex_final} become
\begin{align}
\label{eq:number_ex_lossless}
\left<b^\dagger_m b_l\right>&=\sum_\mu U^*_{m\mu}U_{l\mu}\sinh^2(r_\mu),
\\
\label{eq:corr_ex_lossless}
\left<b_m b_l\right>&= - \sum_\mu U_{m\mu}U_{l\mu}\cosh(r_\mu)\sinh(r_\mu){\rm e}^{i\phi_\mu}.
\end{align}
These are the same results that were derived by Quesada N. \textit{et al.} (see Eqs. (234) - (235) in Ref. \cite{beyondPhotonPairs}). 
\subsection{Lossy single-mode squeezed states}
\label{2modes}
In this subsection we show that for a single lossy mode the coupled differential equations Eqs. \eqref{eq:rdiff} - \eqref{eq:ndiff} give a squeezing amplitude, squeezing phase, and thermal photon number that agrees with previous work on single-mode squeezed thermal states \cite{Seifoory2017SqueezedCavities}. In this case, the Takagi factorization of the nonlinear parameter is trivial since there is only one mode. The matrix $U$ has only a single entry $U_{11} = 1$. The single-mode squeezing operator can be written as
\begin{align}
    \label{eq:squeezeoperator_1mode}
    S(z(t)) &= \exp\left(\frac{1}{2}z_{11}(t)b_1^{\dagger2}-{\rm H.c.}\right),
\end{align}
where  the single squeezing parameter $z_{11}(t)$ is given simply by
\begin{align}
    \label{Takagi_1mode}
    z_{11}(t) = r_1(t) {\rm e}^{i\phi_1(t)}.
\end{align}
 Thus, using Eq. \eqref{eq:Btransform} the quasimode and Schmidt operators are identical $b = B$.  Only keeping the $m=l=\mu=\nu=1$ terms in Eqs. \eqref{eq:rdiff} - \eqref{eq:ndiff} and putting $U_{11}=1$ we obtain
\begin{align}
    \label{eq:rdiff_1mode}
    \Dot{r}_1&=\frac{2|\alpha|^2|\lambda|}{\hbar}-\frac{2\gamma_{ 1}}{2n_1+1}\cosh(r_1)\sinh(r_1),
\\
    \label{eq:phidiff_1mode}
    \Dot{\phi}_1&=2\omega_1,
    \\
    \label{eq:ndiff_1mode}
    \Dot{n}_1&=2\gamma_{1}\left( \sinh^2(r_1)- n_1 \right),
\end{align}
where we have used the phase condition in Eq. \eqref{eq:initialcond_vacuum_phases} again to simplify the equations. These equations are derived for SFWM, but they can be easily adapted for SPDC. The only things that need to change are in Eq. \eqref{eq:rdiff_1mode}, $|\lambda|$ has to be redefined for SPDC, and  $|\alpha|^2$ becomes $|\alpha|$. Therefore, the structure of the equations is unchanged for SPDC. In light of this, these equations are seen to agree with previous work for single-mode squeezed states in a lossy cavity \cite{Seifoory2017SqueezedCavities}.

For a single lossy mode, the expectation values in Eq. \eqref{eq:number_ex_final} and Eq. \eqref{eq:corr_ex_final} become
\begin{align}
\label{eq:number_ex_1mode}
\left<b_1^\dagger b_1\right>&=n_1\cosh(2r_1)+\sinh^2(r_1),
\\
\label{eq:corr_ex_1mode}
\left<b_1 b_1\right>&=-\left(2n_1+1\right)\cosh(r_1)\sinh(r_1).
\end{align}
These are the same equations that were derived in previous work \cite{squeezedthermal}.
\subsection{Lossy two-mode squeezed states}
In this subsection, we show that for two lossy modes the coupled equations give results that agree with our previous work on two-mode squeezed states. For two modes, the nonlinear parameter is a $2\times2$ symmetric matrix and its Takagi factorization  is
\begin{align}
    \label{eq:Takagi_2modes}
    \begin{bmatrix}
    0&G_{12} \\
    G_{12} & 0
    \end{bmatrix}&=U \begin{bmatrix}
    G_{12}&0 \\
    0&G_{12} 
    \end{bmatrix} U^{\rm{T}},
\end{align}
where we neglect generation of squeezed light in single modes by letting  $G_{11} = G_{22} = 0$. The unitary matrix $U$ is given by 
\begin{align}
    \label{eq:TakagiU_2mode}
    U&=\frac{1}{2}\begin{bmatrix}
    1-i&1+i \\
    1+i&1-i
    \end{bmatrix}.
\end{align}
Using Eq. \eqref{eq:takagisqueezingoparam}, the squeezing parameter is given by
\begin{align}
    \label{eq:Takagi_2modes_squeeze}
    \begin{bmatrix}
    0&z_{12}(t) \\
   z_{12}(t) & 0
    \end{bmatrix}&=U \begin{bmatrix}
    r(t){\rm e}^{i\phi(t)}&0 \\
    0&r(t){\rm e}^{i\phi(t)} 
    \end{bmatrix} U^{\rm{T}},
\end{align}
where $z_{12} \equiv r \exp(i\phi)$. There is a single squeezing amplitude ($r(t)$) and squeezing phase ($\phi(t)$), but the thermal photon number for the two modes, $n_1(t)$ and $n_2(t)$, are allowed to be different. This means that the photon loss rates are different for each mode. Putting this $U$ into Eqs. \eqref{eq:rdiff} - \eqref{eq:ndiff} and letting $\lambda = G_{12}$, we obtain
   \begin{align}
   \label{eq:rdiff_2mode}
\dot{r}&=\frac{2|\alpha|^2|\lambda|}{\hbar}-\frac{\cosh(r)\sinh(r)}{1+n_1+n_2}\nonumber
\\
&\times\left[\gamma_1 +\gamma_2 +\left(\gamma_1 -\gamma_2\right)\left(n_2-n_1\right) \right],
\\
\label{eq:phidiff_2mode}
\dot{\phi}&=\left(\omega_1+\omega_2\right),
\\
\label{eq:n1diff_2modes}
\dot{n}_1 &= 2n_1\left[ \gamma_2\sinh^2(r)-\gamma_1\cosh^2(r) \right]+2\gamma_2\sinh^2(r),
    \\
    \label{n2diff_2modes}
    \dot{n}_2 &= 2n_2\left[\gamma_1\sinh^2(r)-\gamma_2\cosh^2(r) \right]+2\gamma_1\sinh^2(r).
 \end{align}
These equations have the same form as those that were derived in our previous work on two-mode squeezed states (see Eqs. (19) - (22) in Ref. \cite{twomodeSL}) generated via SPDC. As mentioned in the previous subsection, we can adapt Eqs. \eqref{eq:rdiff_2mode} - \eqref{n2diff_2modes} for SPDC by replacing $|\alpha|^2$ and $\lambda$ with $|\alpha|$ and the nonlinear parameter for SPDC.
\section{Results for a four-cavity CROW}
\label{results}
In this section we present our results from solving the coupled equations Eqs. \eqref{eq:rdiff} - \eqref{eq:ndiff} for SFWM in a system of four coupled-cavities. The structure is the CROW in Fig. \ref{fig:CROW}, but we take the structure to have only four cavities and we enforce periodic boundary conditions at the two ends of the four-cavity system.  Thus, the quasimodes are Bloch modes, but the allowed Bloch vectors, $k$, are quantized, such that there are only four allowed values. Although such boundary conditions are not physically realizable, we choose this system because it is simple and gives analytic expressions for the complex mode frequencies and the effective nonlinear parameters.  This toy-model is intended to demonstrate how to go about solving the system of equations when there are more than the two modes that have been modeled in previous work \cite{twomodeSL}.

 To obtain our results we use parameters from a CROW  studied in previous work.  The photonic crystal slab that contains the coupled-cavities has a refractive index of $n=3.4$ and a second-order refractive index of $n_2 = 4.5 \times 10^{-18}\,\text{m}^2/\text{W}$, corresponding to Si at telecom wavelengths \cite{chi3silicon}. The photonic crystal is a square lattice with lattice constant, $d$. The slab thickness is $0.8d$ and the air-hole radius is $0.4d$. Using these parameters, with find from finite-difference time domain calculations that each individual cavity contains a single mode at the frequency $\tilde{\omega}_0=(0.305 - i7.71\times 10^{-6})2\pi c/d$, resulting in a quality factor $Q_0= 19800$ \cite{engineeringQ}. The Bloch modes and their frequencies are obtained using a nearest-neighbour tight-binding model, where the individual cavity modes form a basis \cite{pairGenerationCrow}. The computed frequencies of the 4 Bloch modes at $kD = \{-\pi/2,0,\pi/2,\pi\}$ are shown in Tab. \ref{tab:wk}, where $D\equiv 2d$ is the periodicity of the CROW. The expression for the CROW dispersion is approximately given by \cite{Seifoory2019counterpropagatingCV}
 \begin{align}
     \label{eq:crowdispersion}
     \tilde{\omega}_k &\approx \tilde{\omega}_0\left(1 - \tilde{\beta}_1\cos(kD)\right),
 \end{align}
 where $\tilde{\beta}_1 = 9.87\times10^{-3}-i1.97\times10^{-5}$ is the complex coupling parameter between individual nearest-neighbour cavity modes \cite{engineeringQ}. This expression gives the loss dispersion of the CROW as well, it can be obtained by $-2\Im{\tilde{\omega}_k}$. 
 
We let the pump be a \textit{continuous-wave}  with frequency $\omega_P$, modelled by a coherent state with amplitude 
\begin{align}
    \label{eq:cwamplitude}
    \alpha(t) = \left|\alpha\right|{\rm e}^{-i\omega_Pt},
\end{align}
where $|\alpha|^2$ is the average pump photon number. The pump is in the single Bloch mode with wavevector $k_P = \pi/(2D)$ and frequency $\omega_P = 0.305 (2 \pi c/d)$. Signal and idler photons are generated via SFWM into  modes with wavevectors $k_1$ and $k_2$, respectively. The effective nonlinear parameter for this process (see Eq. \eqref{eq:nonlinearparameter}) has been shown to be approximately given by  \cite{pairGenerationCrow} 
\begin{align}
    \label{eq:nonlinearparameterCROW}
    G_{k_1k_2k_Pk_P}\approx G_0 {\rm e}^{-i\Delta k D/2}\frac{{\rm sinc}\left(M\Delta kD/2\right)}{{\rm sinc}\left(\Delta kD/2\right)},
\end{align}
 where $\Delta k \equiv k_1+k_2 - 2k_P$ is the phase mismatch between the different Bloch modes, $M$ is the total number of cavities, and
\begin{align}
\label{eq:G0}
    G_0 \equiv \frac{\hbar^2\omega_0^2\chi^{(3)}_{\rm eff}}{16\epsilon_0(2N+1)V_{\rm eff}},
\end{align}
where $\chi^{(3)}_{\rm eff}$ is the effective nonlinear coefficient, and $V_{\rm eff}$ is the effective mode volume of the individual cavity mode.

\begin{table}
\captionsetup{justification=raggedright}
    \caption{(a) Bloch mode wavevectors $k$ and their complex frequencies $\tilde{\omega}_k$ for a CROW with 4 cavities. Here $D\equiv2d$ is twice the lattice spacing $d$. (b) Diagonal values $\lambda_\mu$ and Schmidt frequencies $\Omega_{\mu \mu}$ for the $\mu$th Schmidt mode. The four Schmidt modes are obtained by taking the Takagi factorization of the nonlinear parameter in Eq. \eqref{eq:nonlinearparameterCROW}}
\begin{subtable}[t]{0.24\textwidth}
\subcaption{}
\renewcommand{\arraystretch}{1.5}
\begin{tabular}{c c}
 \hline
$kD$ &$\tilde{\omega}_k d/(2\pi c)$ \\ \hline \hline 
 $\pi$ & $0.308010  - i1.38\times10^{-5}$ \\ 
$\pm \pi/2$ & $0.305000 - i7.71\times10^{-6}$ \\
0 & $0.301990 - i1.63\times10^{-6}$ \\  \hline
\end{tabular}
 \label{tab:wk}
\end{subtable}
\begin{subtable}[t]{0.23\textwidth}
\subcaption{}
\renewcommand{\arraystretch}{1.5}
\begin{tabular}[t]{c c c }
\hline 
$\mu$ &$\lambda_\mu$  & $\Omega_{\mu \mu} d/(2\pi c)$ \\ \hline \hline
1 & 1.21 & 0.304877 \\ 
2 & 1.16 & 0.304999 \\ 
3 & 0.742 & 0.304953 \\ 
4 & 0.665 & 0.305170 \\ \hline
\end{tabular}
\label{tab:takagi}
\end{subtable}
\end{table}

 Now we perform a Takagi factorization of the nonlinear parameter in Eq. \eqref{eq:nonlinearparameterCROW}. The resulting diagonal values $\lambda_\mu$ and Schmidt mode frequencies $\Omega_{\mu \mu}$ (see Eq. \eqref{transformed_freq}) are shown in Tab. \ref{tab:takagi}. The frequencies $\Omega_{\mu \mu}$ are all within $0.06\%$ of the pump frequency $\omega_P$. This is due to the dispersion of the Bloch mode frequencies having a small bandwidth and choosing the pump to be at the center of the band. This will not be the case for a general structure. As we show below, however, the Schmidt modes with frequencies that are nearly on resonance with the pump have larger squeezing amplitudes than those that are off resonance. 

To obtain a particular solution, we initially let the system be in the vacuum state by setting all the squeezing amplitudes and thermal photon numbers equal to zero. In Appendix \ref{solving} we discuss how to solve the coupled equations starting from the vacuum, particularly how to choose the initial squeezing phase. At time $t=0$ the pump is put into the Bloch mode $k_P=\pi/(2D)$ and then it generates signal and idler photons in the Bloch modes $k_1$ and $k_2$, respectively. 
Using these initial conditions, we solve Eqs. \eqref{eq:rdiff} - \eqref{eq:ndiff} numerically using a Runge-Kutta method (such as the ode45 function in MATLAB). For 4 modes it takes on the order of 10 seconds to solve the equations using a standard PC.

In what follows, we use $t_{\rm c}$ as our unit of time, where $t_{\rm c}$ is defined as the time it takes a pulse of light with group velocity $v$ to propagate the total distance, $L$, of the coupled-cavity structure:
\begin{align}
    \label{eq:tcrow}
    t_{\rm c} \equiv \frac{L}{v}.
\end{align}
For 4 cavities with a lattice constant of $d=480$nm, the length of the structure is $L= 2.9 \mu$m. If the pulse is centered at $kD=\pi/2$, and assuming linear dispersion in Eq. \eqref{eq:crowdispersion}, then the group velocity of the pulse is approximately $v = c/26.6$. Therefore, we obtain $t_{\rm c} = 0.25$ps. Additionally, we define the pumping strength dimensionless parameter,
\begin{align}
    \label{eq:pumpstrength}
    g\equiv \frac{4G_0\left|\alpha\right|^2t_{\rm c}}{\hbar},
\end{align}
 which scales the squeezing amplitude in all the Schmidt modes. In all that follows, we set $g=1/12$. This value of $g$ can be achieved with an average pump photon number of $|\alpha|^2 = 4.6\times 10^{7}$ and an effective mode volume of $V_{\rm eff} = 3 (\mu{\rm m})^3$ for the cavity mode with frequency $\tilde{\omega}_0$. It gives squeezing amplitudes in the Schmidt modes that are on the order of one, while keeping the thermal photon number many orders of magnitude below the number of photons in the pump.

\begin{figure}[h!]
     \centering
     \begin{subfigure}[b]{0.48\textwidth}
         \centering
         \includegraphics[width=\textwidth]{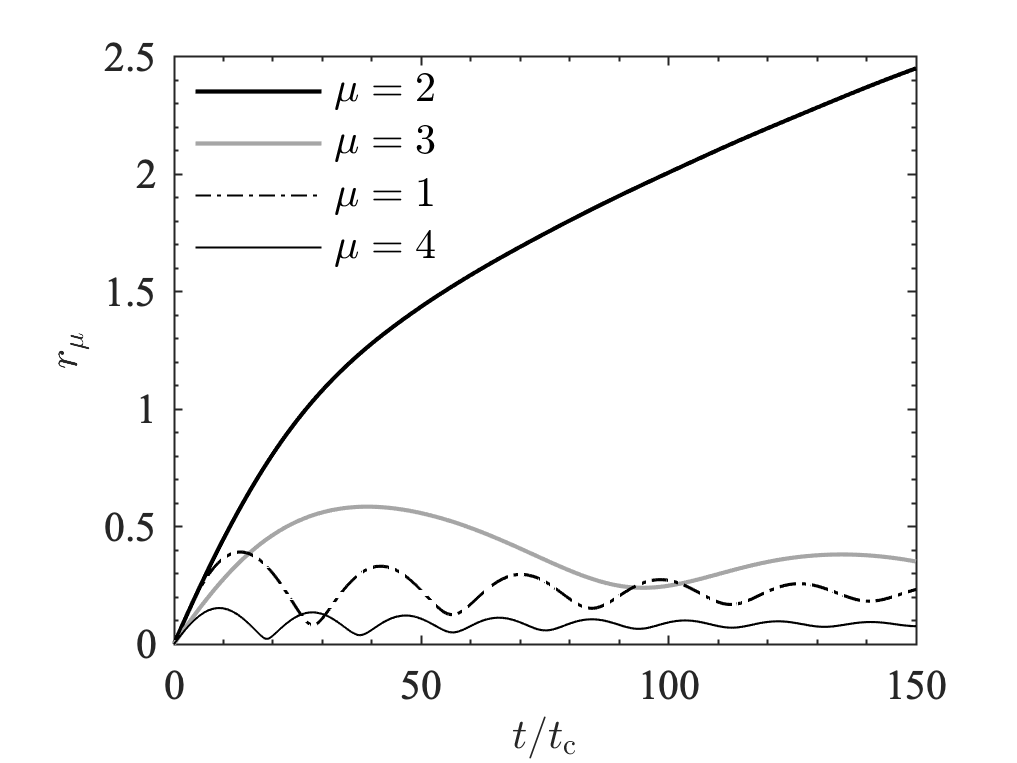}
         \caption{}
         \label{fig:rcrow}
     \end{subfigure}
     \begin{subfigure}[b]{0.48\textwidth}
         \centering
         \includegraphics[width=\textwidth]{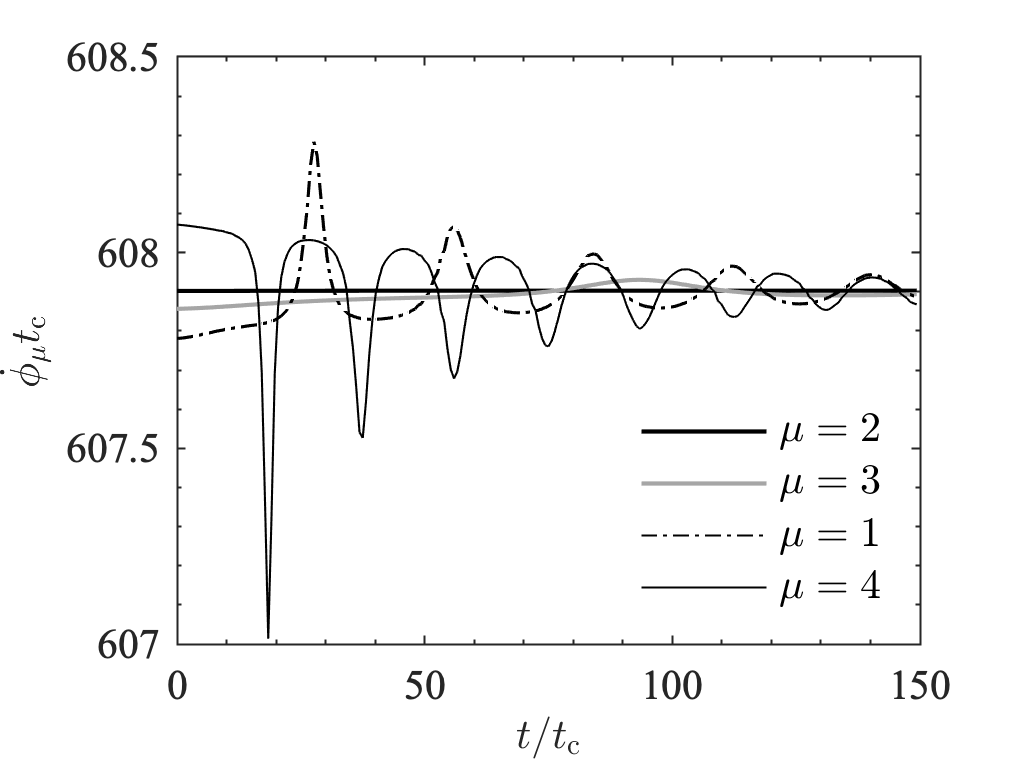}
         \caption{}
         \label{fig:phicrow}
     \end{subfigure}
          \caption{(a) Squeezing amplitudes $r_\mu$ and (b) derivatives of the squeezing phases $\dot{\phi}_\mu$ in the four Schmidt modes of four coupled-cavities. Here $t_{\rm c}=0.25$ps is the time for a light pulse to cross the length of the structure.}
\end{figure}

We start by considering the Schmidt mode squeezing amplitudes in Fig. \ref{fig:rcrow}. At $t=0$ the pump is turned on and the $r_\mu$ initially increase linearly with time, where it can be shown using Eq. \eqref{eq:rdiff} that the slope is approximately given by $\dot{r}_\mu\approx g|\lambda_\mu|/2$ (see also Eq. \eqref{eq:r_t=dt}). This is because for short times the thermal noise and detuning $|\Omega_{\mu\mu}-\omega_P|$ can be neglected. The Schmidt modes that are detuned from the pump will have oscillations in their $r_\mu$ and an $r_\mu$ that is smaller than that of the modes that are on resonance with the pump. The oscillations exist for Schmidt modes that have a detuning from the pump frequency that satisfies $|\Omega_{\mu\mu} - \omega_P|t_{\rm c}>g|\lambda_\mu|/2$. Thus, these oscillations will disappear if we increase the pumping strength $g$. If, however, $g$ is small, then the period of the oscillations are approximately given by $\pi(|\Omega_{\mu\mu}-\omega_P|t_{\rm c})^{-1}$. For example, using this expression, the squeezing amplitude in the $\mu=1$ and $\mu=4$ Schmidt modes have periods of approximately $26 t_{\rm c}$ and $19 t_{\rm c}$, respectively. As time increases, the amplitude of the oscillations are dampened due to the intrinsic losses of the Bloch modes. 

Next, we consider the derivative of the Schmidt mode squeezing phases, $\dot{\phi}_\mu$, in Fig. \ref{fig:phicrow}. Initially the derivative is approximately constant, corresponding to the phases increasing linearly with time. In  Appendix \ref{solving} we show that for short times $\Dot{\phi}_\mu(t) \approx \Omega_{\mu \mu} + \omega_P$. The modes $\mu=2$ and $\mu=3$ that are close to resonance with the pump have an approximately constant phase derivative approaching $2\omega_P$. The detuned modes $\mu=1$ and $\mu=4$ have peaks in their phase derivatives whenever the squeezing amplitude is close to zero. This is because in Eq. \eqref{eq:phidiff} the term proportional to $1/\tanh(2r_\mu)$ is large when $r_\mu\ll1$, which causes $\Dot{\phi}_\mu$ to increase.

\begin{figure}[h!]
     \centering
     \begin{subfigure}[b]{0.48\textwidth}
         \centering
         \includegraphics[width=\textwidth]{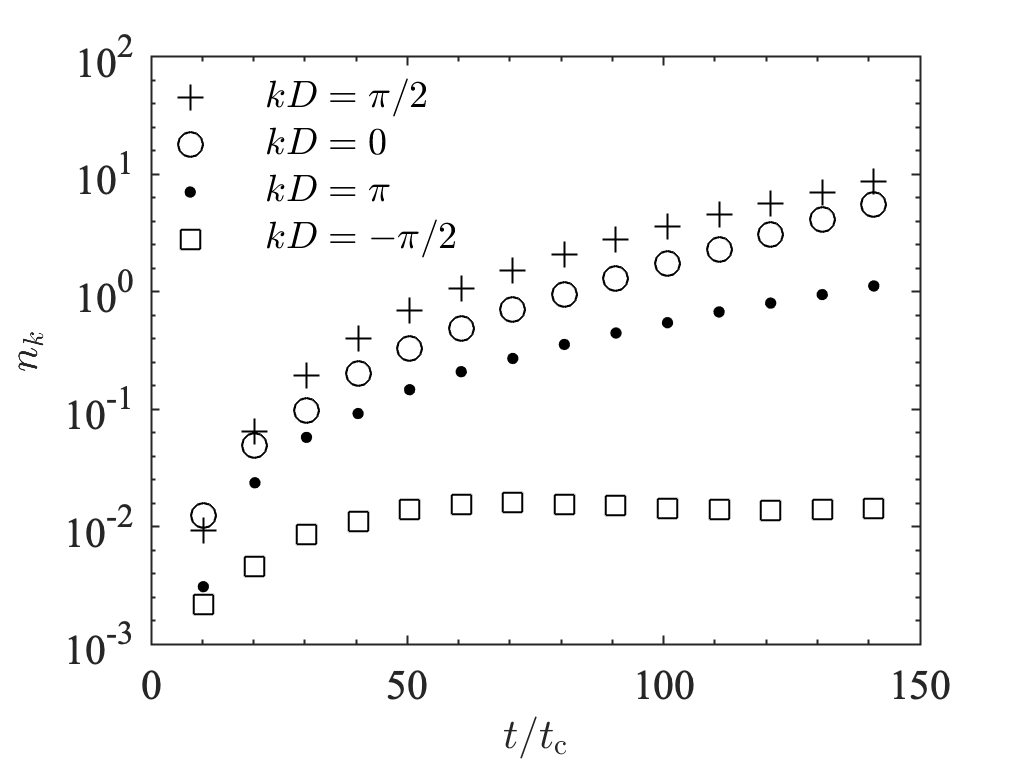}
         \caption{}
         \label{fig:nthcrow}
     \end{subfigure}
     \begin{subfigure}[b]{0.48\textwidth}
         \centering
         \includegraphics[width=\textwidth]{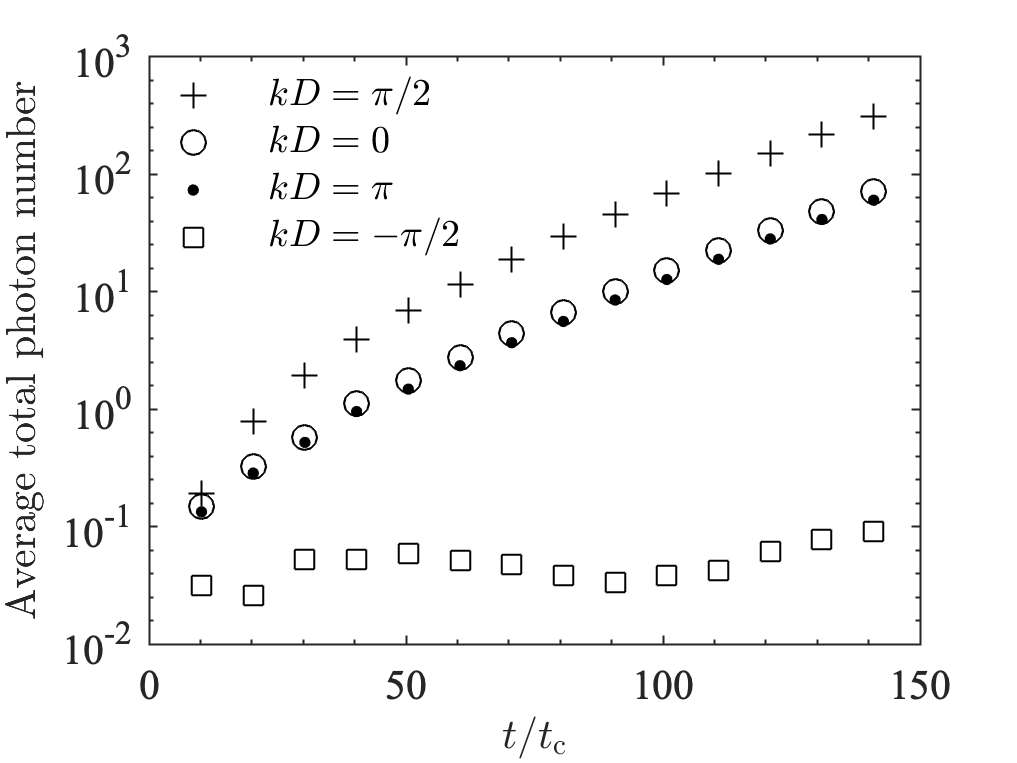}
         \caption{}
         \label{fig:ntotcrow}
     \end{subfigure}
          \caption{(a) Thermal photon number $n_k$ and (b) average total photon number in the Bloch modes of four coupled-cavities. }
\end{figure}

Now, we present the average thermal photon numbers in the Bloch modes, $n_k$, in Fig. \ref{fig:nthcrow}. At short times $t<10t_{\rm c}$ the $n_k$ are negligible. The $n_k$ do not go above approximately $10$ for the pumping strength $g=1/12$. The majority of the thermal noise is generated in the Bloch modes with positive wavevectors. The $n_k$ in the Bloch mode with $kD = -\pi/2$ is only approximately $0.01$. 

In Fig. \ref{fig:ntotcrow} we calculate the average total photon number using Eq. \eqref{eq:corr_ex}. There are on the order of approximately $10^3$ total photons generated by the nonlinear process. The majority are generated in the forward-propagating Bloch mode, $kD=\pi/2$ and $kD=\pi$ and $k=0$. Only approximately $0.1$ photons, on average, are generated in the backward-propagating Bloch mode, $kD = -\pi/2$. 
\section{Conclusion}
\label{Conclusion}
The key finding in this paper is that the solution to the Lindblad master equation for a set of lossy quasimodes is the density operator for a multimode squeezed thermal state (see Eq. \eqref{eq:MSTS}). In order to prove this we introduce an orthogonal Schmidt basis via the Takagi factorization in Sec. \ref{takagi} to diagonalize the nonlinear Hamiltonian and squeezing operator. The main result is a set of coupled first-order differential equations that the squeezing amplitude, squeezing phase, and thermal photon numbers must obey in order for the multimode squeezed thermal state to be a solution (see Eqs. \eqref{eq:rdiff} - \eqref{eq:ndiff}). 

In order to derive the solution, we make the undepleted pump approximation on the nonlinear Hamiltonian and assume that the nonlinear parameter is essentially the same for all pump modes in the pump bandwidth and neglect its dependence on the pump frequencies (see Sec. \ref{pumpconsiderations}). The latter assumption is valid for a  pump pulse in a waveguide or free space as long as the pulse is long in duration. 

Our theory is applicable to the orthogonal set of discrete lossy modes in structures, such as ring resonators, coupled-ring resonators, CROWs, and high-Q cavities coupled to a waveguide in a photonic crystal (see Fig. \ref{fig:structures}).

 Our results are consistent with previous work done on single-mode and two-mode squeezed thermal states (see Refs. \cite{Seifoory2017SqueezedCavities} and \cite{twomodeSL}) as presented in Sec. \ref{lmitingcases}. Also in the limit where the quasimodes become lossless, our solution reduces to a multimode squeezed vacuum state, and it agrees with other work \cite{beyondPhotonPairs} that used a similar approach also using a Takagi factorization.
 
To our knowledge, this is the first time the analytic solution for the quantum state of light generated via nonlinear processes in $M>2$ lossy modes has been derived. Not only is the solution of theoretical importance, it can greatly reduce the number of coupled differential equations required to solve for the density operator. One alternative way to determine the density operator numerically, is by calculating its matrix elements using Fock states. The number of possible states for $N$ photons in $M$ quasimodes is $\sum_{j=0}^N\binom{j+M-1}{j} = (M+N)!/(M!N!)$ \cite{numericalLindblad}, where $\binom{n}{k} = n!/[k!(n-k)!]$ is the binomial coefficient. In order to find the matrix elements, the number of coupled equations one has to solve is the square of this. However, using our results one can determine the density operator by solving only $3M$ coupled first-order differential equations that are independent of $N$. For example if $N=20$ and $M=4$ there are $10 626$ basis states and thus  $(10 626)^2$ coupled equations. But our theory would only require solving $12$ coupled equations.
 
The expressions we derive for the correlation variances (see Sec. \ref{expectation}) only contain double sums over the Schmidt modes and  do not require one to solve any additional coupled equations. Thus, the differential equations only need to be solved once and then any same-time correlation variances of interest can be quickly calculated. 
 
   We believe that our new solution to this important problem will make it more feasible to study large multimode lossy structures and to optimize them for a wide variety of quantum information applications. In future work, we will apply our theory to more physically realistic structures, such as many-cavity CROWs and ring resonators. We then hope to show that the generated multimode squeezed thermal state is an approximate Gaussian cluster state \cite{anyclusterstate,graphicalCalcGuassianStates}. To do this, one needs to determine under what conditions on the nonlinear parameter and/or the squeezing matrix, does the state satisfy nullifier equations. If it is possible to determine these conditions for the multimode squeezed thermal state we derived, then it can be used as a cluster state resource for quantum computations \cite{wuCVchip2020}. 

\section*{Acknowledgements}
This work was supported by Queen’s University and the Natural Sciences and Engineering Research Council of Canada (NSERC).

\bibliography{apssamp}

\providecommand{\noopsort}[1]{}\providecommand{\singleletter}[1]{#1}%
\begin{thebibliography}{35}%
\makeatletter
\providecommand \@ifxundefined [1]{%
 \@ifx{#1\undefined}
}%
\providecommand \@ifnum [1]{%
 \ifnum #1\expandafter \@firstoftwo
 \else \expandafter \@secondoftwo
 \fi
}%
\providecommand \@ifx [1]{%
 \ifx #1\expandafter \@firstoftwo
 \else \expandafter \@secondoftwo
 \fi
}%
\providecommand \natexlab [1]{#1}%
\providecommand \enquote  [1]{``#1''}%
\providecommand \bibnamefont  [1]{#1}%
\providecommand \bibfnamefont [1]{#1}%
\providecommand \citenamefont [1]{#1}%
\providecommand \href@noop [0]{\@secondoftwo}%
\providecommand \href [0]{\begingroup \@sanitize@url \@href}%
\providecommand \@href[1]{\@@startlink{#1}\@@href}%
\providecommand \@@href[1]{\endgroup#1\@@endlink}%
\providecommand \@sanitize@url [0]{\catcode `\\12\catcode `\$12\catcode
  `\&12\catcode `\#12\catcode `\^12\catcode `\_12\catcode `\%12\relax}%
\providecommand \@@startlink[1]{}%
\providecommand \@@endlink[0]{}%
\providecommand \url  [0]{\begingroup\@sanitize@url \@url }%
\providecommand \@url [1]{\endgroup\@href {#1}{\urlprefix }}%
\providecommand \urlprefix  [0]{URL }%
\providecommand \Eprint [0]{\href }%
\providecommand \doibase [0]{https://doi.org/}%
\providecommand \selectlanguage [0]{\@gobble}%
\providecommand \bibinfo  [0]{\@secondoftwo}%
\providecommand \bibfield  [0]{\@secondoftwo}%
\providecommand \translation [1]{[#1]}%
\providecommand \BibitemOpen [0]{}%
\providecommand \bibitemStop [0]{}%
\providecommand \bibitemNoStop [0]{.\EOS\space}%
\providecommand \EOS [0]{\spacefactor3000\relax}%
\providecommand \BibitemShut  [1]{\csname bibitem#1\endcsname}%
\let\auto@bib@innerbib\@empty
\bibitem [{\citenamefont {Hamilton}\ \emph {et~al.}(2017)\citenamefont
  {Hamilton}, \citenamefont {Kruse}, \citenamefont {Sansoni}, \citenamefont
  {Barkhofen}, \citenamefont {Silberhorn},\ and\ \citenamefont
  {Jex}}]{GaussianBosonSampling}%
  \BibitemOpen
  \bibfield  {author} {\bibinfo {author} {\bibfnamefont {C.~S.}\ \bibnamefont
  {Hamilton}}, \bibinfo {author} {\bibfnamefont {R.}~\bibnamefont {Kruse}},
  \bibinfo {author} {\bibfnamefont {L.}~\bibnamefont {Sansoni}}, \bibinfo
  {author} {\bibfnamefont {S.}~\bibnamefont {Barkhofen}}, \bibinfo {author}
  {\bibfnamefont {C.}~\bibnamefont {Silberhorn}},\ and\ \bibinfo {author}
  {\bibfnamefont {I.}~\bibnamefont {Jex}},\ }\bibfield  {title} {\bibinfo
  {title} {Gaussian boson sampling},\ }\href@noop {} {\bibfield  {journal}
  {\bibinfo  {journal} {Phys. Rev. Lett.}\ }\textbf {\bibinfo {volume} {119}},\
  \bibinfo {pages} {170501} (\bibinfo {year} {2017})}\BibitemShut {NoStop}%
\bibitem [{\citenamefont {Kruse}\ \emph {et~al.}(2019)\citenamefont {Kruse},
  \citenamefont {Hamilton}, \citenamefont {Sansoni}, \citenamefont {Barkhofen},
  \citenamefont {Silberhorn},\ and\ \citenamefont
  {Jex}}]{DetailedGaussianBosonSampling}%
  \BibitemOpen
  \bibfield  {author} {\bibinfo {author} {\bibfnamefont {R.}~\bibnamefont
  {Kruse}}, \bibinfo {author} {\bibfnamefont {C.~S.}\ \bibnamefont {Hamilton}},
  \bibinfo {author} {\bibfnamefont {L.}~\bibnamefont {Sansoni}}, \bibinfo
  {author} {\bibfnamefont {S.}~\bibnamefont {Barkhofen}}, \bibinfo {author}
  {\bibfnamefont {C.}~\bibnamefont {Silberhorn}},\ and\ \bibinfo {author}
  {\bibfnamefont {I.}~\bibnamefont {Jex}},\ }\bibfield  {title} {\bibinfo
  {title} {Detailed study of gaussian boson sampling},\ }\href@noop {}
  {\bibfield  {journal} {\bibinfo  {journal} {Phys. Rev. A}\ }\textbf {\bibinfo
  {volume} {100}},\ \bibinfo {pages} {032326} (\bibinfo {year}
  {2019})}\BibitemShut {NoStop}%
\bibitem [{\citenamefont {Braunstein}\ and\ \citenamefont {van
  Loock}(2005)}]{Braunstein2005QuantumVariables}%
  \BibitemOpen
  \bibfield  {author} {\bibinfo {author} {\bibfnamefont {S.~L.}\ \bibnamefont
  {Braunstein}}\ and\ \bibinfo {author} {\bibfnamefont {P.}~\bibnamefont {van
  Loock}},\ }\bibfield  {title} {\bibinfo {title} {{Quantum information with
  continuous variables}},\ }\href@noop {} {\bibfield  {journal} {\bibinfo
  {journal} {Rev. Mod. Phys.}\ }\textbf {\bibinfo {volume} {77}},\ \bibinfo
  {pages} {513} (\bibinfo {year} {2005})}\BibitemShut {NoStop}%
\bibitem [{\citenamefont {Takeda}\ and\ \citenamefont
  {Furusawa}(2019)}]{Takeda2019TowardComputing}%
  \BibitemOpen
  \bibfield  {author} {\bibinfo {author} {\bibfnamefont {S.}~\bibnamefont
  {Takeda}}\ and\ \bibinfo {author} {\bibfnamefont {A.}~\bibnamefont
  {Furusawa}},\ }\bibfield  {title} {\bibinfo {title} {{Toward large-scale
  fault-tolerant universal photonic quantum computing}},\ }\href@noop {}
  {\bibfield  {journal} {\bibinfo  {journal} {APL Photonics}\ }\textbf
  {\bibinfo {volume} {4}},\ \bibinfo {pages} {060902} (\bibinfo {year}
  {2019})}\BibitemShut {NoStop}%
\bibitem [{\citenamefont {Pfister}(2020)}]{CVquantumComputing}%
  \BibitemOpen
  \bibfield  {author} {\bibinfo {author} {\bibfnamefont {O.}~\bibnamefont
  {Pfister}},\ }\bibfield  {title} {\bibinfo {title} {{Continuous-variable
  quantum computing in the quantum optical frequency comb}},\ }\href@noop {}
  {\bibfield  {journal} {\bibinfo  {journal} {J. Phys. B: At. Mol. Opt. Phys.}\
  }\textbf {\bibinfo {volume} {53}},\ \bibinfo {pages} {012001} (\bibinfo
  {year} {2020})}\BibitemShut {NoStop}%
\bibitem [{\citenamefont {Zhang}\ and\ \citenamefont
  {Braunstein}(2006)}]{CVclusterstateBraunstein}%
  \BibitemOpen
  \bibfield  {author} {\bibinfo {author} {\bibfnamefont {J.}~\bibnamefont
  {Zhang}}\ and\ \bibinfo {author} {\bibfnamefont {S.~L.}\ \bibnamefont
  {Braunstein}},\ }\bibfield  {title} {\bibinfo {title} {Continuous-variable
  gaussian analog of cluster states},\ }\href@noop {} {\bibfield  {journal}
  {\bibinfo  {journal} {Phys. Rev. A}\ }\textbf {\bibinfo {volume} {73}},\
  \bibinfo {pages} {032318} (\bibinfo {year} {2006})}\BibitemShut {NoStop}%
\bibitem [{\citenamefont {Menicucci}\ \emph {et~al.}(2011)\citenamefont
  {Menicucci}, \citenamefont {Flammia},\ and\ \citenamefont {van
  Loock}}]{graphicalCalcGuassianStates}%
  \BibitemOpen
  \bibfield  {author} {\bibinfo {author} {\bibfnamefont {N.~C.}\ \bibnamefont
  {Menicucci}}, \bibinfo {author} {\bibfnamefont {S.~T.}\ \bibnamefont
  {Flammia}},\ and\ \bibinfo {author} {\bibfnamefont {P.}~\bibnamefont {van
  Loock}},\ }\bibfield  {title} {\bibinfo {title} {Graphical calculus for
  gaussian pure states},\ }\href@noop {} {\bibfield  {journal} {\bibinfo
  {journal} {Phys. Rev. A}\ }\textbf {\bibinfo {volume} {83}},\ \bibinfo
  {pages} {042335} (\bibinfo {year} {2011})}\BibitemShut {NoStop}%
\bibitem [{\citenamefont {Wu}\ \emph {et~al.}(2020)\citenamefont {Wu},
  \citenamefont {Alexander}, \citenamefont {Liu},\ and\ \citenamefont
  {Zhang}}]{wuCVchip2020}%
  \BibitemOpen
  \bibfield  {author} {\bibinfo {author} {\bibfnamefont {B.~H.}\ \bibnamefont
  {Wu}}, \bibinfo {author} {\bibfnamefont {R.~N.}\ \bibnamefont {Alexander}},
  \bibinfo {author} {\bibfnamefont {S.}~\bibnamefont {Liu}},\ and\ \bibinfo
  {author} {\bibfnamefont {Z.}~\bibnamefont {Zhang}},\ }\bibfield  {title}
  {\bibinfo {title} {Quantum computing with multidimensional
  continuous-variable cluster states in a scalable photonic platform},\
  }\href@noop {} {\bibfield  {journal} {\bibinfo  {journal} {Phys. Rev. Res.}\
  }\textbf {\bibinfo {volume} {2}},\ \bibinfo {pages} {023138} (\bibinfo {year}
  {2020})}\BibitemShut {NoStop}%
\bibitem [{\citenamefont {Vernon}\ and\ \citenamefont
  {Sipe}(2015)}]{strongquantumopticsringresonator}%
  \BibitemOpen
  \bibfield  {author} {\bibinfo {author} {\bibfnamefont {Z.}~\bibnamefont
  {Vernon}}\ and\ \bibinfo {author} {\bibfnamefont {J.~E.}\ \bibnamefont
  {Sipe}},\ }\bibfield  {title} {\bibinfo {title} {Strongly driven nonlinear
  quantum optics in microring resonators},\ }\href@noop {} {\bibfield
  {journal} {\bibinfo  {journal} {Phys. Rev. A}\ }\textbf {\bibinfo {volume}
  {92}},\ \bibinfo {pages} {033840} (\bibinfo {year} {2015})}\BibitemShut
  {NoStop}%
\bibitem [{\citenamefont {Yang}\ \emph {et~al.}(2008)\citenamefont {Yang},
  \citenamefont {Liscidini},\ and\ \citenamefont {Sipe}}]{backwardHeisenberg}%
  \BibitemOpen
  \bibfield  {author} {\bibinfo {author} {\bibfnamefont {Z.}~\bibnamefont
  {Yang}}, \bibinfo {author} {\bibfnamefont {M.}~\bibnamefont {Liscidini}},\
  and\ \bibinfo {author} {\bibfnamefont {J.~E.}\ \bibnamefont {Sipe}},\
  }\bibfield  {title} {\bibinfo {title} {Spontaneous parametric down-conversion
  in waveguides: A backward heisenberg picture approach},\ }\href@noop {}
  {\bibfield  {journal} {\bibinfo  {journal} {Phys. Rev. A}\ }\textbf {\bibinfo
  {volume} {77}},\ \bibinfo {pages} {033808} (\bibinfo {year}
  {2008})}\BibitemShut {NoStop}%
\bibitem [{\citenamefont {Sharapova}\ \emph {et~al.}(2020)\citenamefont
  {Sharapova}, \citenamefont {Frascella}, \citenamefont {Riabinin},
  \citenamefont {P\'erez}, \citenamefont {Tikhonova}, \citenamefont {Lemieux},
  \citenamefont {Boyd}, \citenamefont {Leuchs},\ and\ \citenamefont
  {Chekhova}}]{brightsqueezedvacuum}%
  \BibitemOpen
  \bibfield  {author} {\bibinfo {author} {\bibfnamefont {P.~R.}\ \bibnamefont
  {Sharapova}}, \bibinfo {author} {\bibfnamefont {G.}~\bibnamefont
  {Frascella}}, \bibinfo {author} {\bibfnamefont {M.}~\bibnamefont {Riabinin}},
  \bibinfo {author} {\bibfnamefont {A.~M.}\ \bibnamefont {P\'erez}}, \bibinfo
  {author} {\bibfnamefont {O.~V.}\ \bibnamefont {Tikhonova}}, \bibinfo {author}
  {\bibfnamefont {S.}~\bibnamefont {Lemieux}}, \bibinfo {author} {\bibfnamefont
  {R.~W.}\ \bibnamefont {Boyd}}, \bibinfo {author} {\bibfnamefont
  {G.}~\bibnamefont {Leuchs}},\ and\ \bibinfo {author} {\bibfnamefont {M.~V.}\
  \bibnamefont {Chekhova}},\ }\bibfield  {title} {\bibinfo {title} {Properties
  of bright squeezed vacuum at increasing brightness},\ }\href@noop {}
  {\bibfield  {journal} {\bibinfo  {journal} {Phys. Rev. Res.}\ }\textbf
  {\bibinfo {volume} {2}},\ \bibinfo {pages} {013371} (\bibinfo {year}
  {2020})}\BibitemShut {NoStop}%
\bibitem [{\citenamefont {Seifoory}\ \emph {et~al.}(2019)\citenamefont
  {Seifoory}, \citenamefont {Helt}, \citenamefont {Sipe},\ and\ \citenamefont
  {Dignam}}]{Seifoory2019counterpropagatingCV}%
  \BibitemOpen
  \bibfield  {author} {\bibinfo {author} {\bibfnamefont {H.}~\bibnamefont
  {Seifoory}}, \bibinfo {author} {\bibfnamefont {L.~G.}\ \bibnamefont {Helt}},
  \bibinfo {author} {\bibfnamefont {J.~E.}\ \bibnamefont {Sipe}},\ and\
  \bibinfo {author} {\bibfnamefont {M.~M.}\ \bibnamefont {Dignam}},\ }\bibfield
   {title} {\bibinfo {title} {Counterpropagating continuous-variable entangled
  states in lossy coupled-cavity optical waveguides},\ }\href@noop {}
  {\bibfield  {journal} {\bibinfo  {journal} {Phys. Rev. A}\ }\textbf {\bibinfo
  {volume} {100}},\ \bibinfo {pages} {033839} (\bibinfo {year}
  {2019})}\BibitemShut {NoStop}%
\bibitem [{\citenamefont {Dezfouli}\ and\ \citenamefont
  {Dignam}(2017)}]{pairGenerationCrow}%
  \BibitemOpen
  \bibfield  {author} {\bibinfo {author} {\bibfnamefont {M.~K.}\ \bibnamefont
  {Dezfouli}}\ and\ \bibinfo {author} {\bibfnamefont {M.~M.}\ \bibnamefont
  {Dignam}},\ }\bibfield  {title} {\bibinfo {title} {Photon-pair generation in
  lossy coupled-resonator optical waveguides via spontaneous four-wave
  mixing},\ }\href@noop {} {\bibfield  {journal} {\bibinfo  {journal} {Phys.
  Rev. A}\ }\textbf {\bibinfo {volume} {95}},\ \bibinfo {pages} {033815}
  (\bibinfo {year} {2017})}\BibitemShut {NoStop}%
\bibitem [{\citenamefont {Helt}\ and\ \citenamefont
  {Quesada}(2020)}]{degenerateSqueezingUnified}%
  \BibitemOpen
  \bibfield  {author} {\bibinfo {author} {\bibfnamefont {L.}~\bibnamefont
  {Helt}}\ and\ \bibinfo {author} {\bibfnamefont {N.}~\bibnamefont {Quesada}},\
  }\bibfield  {title} {\bibinfo {title} {{Degenerate squeezing in waveguides: A
  unified theoretical approach}},\ }\href@noop {} {\bibfield  {journal}
  {\bibinfo  {journal} {J. Phys. Photonics}\ }\textbf {\bibinfo {volume} {2}},\
  \bibinfo {pages} {035001} (\bibinfo {year} {2020})}\BibitemShut {NoStop}%
\bibitem [{\citenamefont {Gregg}\ and\ \citenamefont
  {Richter}(2016)}]{numericalLindblad}%
  \BibitemOpen
  \bibfield  {author} {\bibinfo {author} {\bibfnamefont {M.}~\bibnamefont
  {Gregg}}\ and\ \bibinfo {author} {\bibfnamefont {M.}~\bibnamefont
  {Richter}},\ }\bibfield  {title} {\bibinfo {title} {Efficient and exact
  numerical approach for many multi-level systems in open system {CQED}},\
  }\href@noop {} {\bibfield  {journal} {\bibinfo  {journal} {New J. Phys.}\
  }\textbf {\bibinfo {volume} {18}},\ \bibinfo {pages} {043037} (\bibinfo
  {year} {2016})}\BibitemShut {NoStop}%
\bibitem [{\citenamefont {Quesada}\ \emph {et~al.}()\citenamefont {Quesada},
  \citenamefont {Helt}, \citenamefont {Menotti}, \citenamefont {Liscidini},\
  and\ \citenamefont {Sipe}}]{beyondPhotonPairs}%
  \BibitemOpen
  \bibfield  {author} {\bibinfo {author} {\bibfnamefont {N.}~\bibnamefont
  {Quesada}}, \bibinfo {author} {\bibfnamefont {L.~G.}\ \bibnamefont {Helt}},
  \bibinfo {author} {\bibfnamefont {M.}~\bibnamefont {Menotti}}, \bibinfo
  {author} {\bibfnamefont {M.}~\bibnamefont {Liscidini}},\ and\ \bibinfo
  {author} {\bibfnamefont {J.~E.}\ \bibnamefont {Sipe}},\ }\bibfield  {title}
  {\bibinfo {title} {{Beyond photon pairs: Nonlinear quantum photonics in the
  high-gain regime}},\ }\href@noop {} {\bibinfo  {journal} {arXiv:2110.04340}\
  }\BibitemShut {NoStop}%
\bibitem [{\citenamefont {Kim}\ \emph {et~al.}(1989)\citenamefont {Kim},
  \citenamefont {de~Oliveira},\ and\ \citenamefont {Knight}}]{squeezedthermal}%
  \BibitemOpen
\bibfield  {journal} {  }\bibfield  {author} {\bibinfo {author} {\bibfnamefont
  {M.~S.}\ \bibnamefont {Kim}}, \bibinfo {author} {\bibfnamefont {F.~A.~M.}\
  \bibnamefont {de~Oliveira}},\ and\ \bibinfo {author} {\bibfnamefont {P.~L.}\
  \bibnamefont {Knight}},\ }\bibfield  {title} {\bibinfo {title} {Properties of
  squeezed number states and squeezed thermal states},\ }\href@noop {}
  {\bibfield  {journal} {\bibinfo  {journal} {Phys. Rev. A}\ }\textbf {\bibinfo
  {volume} {40}},\ \bibinfo {pages} {2494} (\bibinfo {year}
  {1989})}\BibitemShut {NoStop}%
\bibitem [{\citenamefont {Seifoory}\ \emph {et~al.}(2017)\citenamefont
  {Seifoory}, \citenamefont {Doutre}, \citenamefont {Dignam},\ and\
  \citenamefont {Sipe}}]{Seifoory2017SqueezedCavities}%
  \BibitemOpen
  \bibfield  {author} {\bibinfo {author} {\bibfnamefont {H.}~\bibnamefont
  {Seifoory}}, \bibinfo {author} {\bibfnamefont {S.}~\bibnamefont {Doutre}},
  \bibinfo {author} {\bibfnamefont {M.}~\bibnamefont {Dignam}},\ and\ \bibinfo
  {author} {\bibfnamefont {J.~E.}\ \bibnamefont {Sipe}},\ }\bibfield  {title}
  {\bibinfo {title} {Squeezed thermal states: the result of parametric down
  conversion in lossy cavities},\ }\href@noop {} {\bibfield  {journal}
  {\bibinfo  {journal} {J. Opt. Soc. Am. B}\ }\textbf {\bibinfo {volume}
  {34}},\ \bibinfo {pages} {1587} (\bibinfo {year} {2017})}\BibitemShut
  {NoStop}%
\bibitem [{\citenamefont {Vendromin}\ and\ \citenamefont
  {Dignam}(2021)}]{twomodeSL}%
  \BibitemOpen
  \bibfield  {author} {\bibinfo {author} {\bibfnamefont {C.}~\bibnamefont
  {Vendromin}}\ and\ \bibinfo {author} {\bibfnamefont {M.~M.}\ \bibnamefont
  {Dignam}},\ }\bibfield  {title} {\bibinfo {title} {Continuous-variable
  entanglement in a two-mode lossy cavity: An analytic solution},\ }\href@noop
  {} {\bibfield  {journal} {\bibinfo  {journal} {Phys. Rev. A}\ }\textbf
  {\bibinfo {volume} {103}},\ \bibinfo {pages} {022418} (\bibinfo {year}
  {2021})}\BibitemShut {NoStop}%
\bibitem [{\citenamefont {Fussell}\ and\ \citenamefont
  {Dignam}(2008)}]{quasimodeapproach}%
  \BibitemOpen
  \bibfield  {author} {\bibinfo {author} {\bibfnamefont {D.~P.}\ \bibnamefont
  {Fussell}}\ and\ \bibinfo {author} {\bibfnamefont {M.~M.}\ \bibnamefont
  {Dignam}},\ }\bibfield  {title} {\bibinfo {title} {Quasimode-projection
  approach to quantum-dot–photon interactions in photonic-crystal-slab
  coupled-cavity systems},\ }\href@noop {} {\bibfield  {journal} {\bibinfo
  {journal} {Phys. Rev. A}\ }\textbf {\bibinfo {volume} {77}},\ \bibinfo
  {pages} {053805} (\bibinfo {year} {2008})}\BibitemShut {NoStop}%
\bibitem [{\citenamefont {Dignam}\ \emph {et~al.}(2006)\citenamefont {Dignam},
  \citenamefont {Fussell}, \citenamefont {Steel}, \citenamefont {Martijn~de
  Sterke},\ and\ \citenamefont {McPhedran}}]{spontaneousEmissionSuppression}%
  \BibitemOpen
  \bibfield  {author} {\bibinfo {author} {\bibfnamefont {M.~M.}\ \bibnamefont
  {Dignam}}, \bibinfo {author} {\bibfnamefont {D.~P.}\ \bibnamefont {Fussell}},
  \bibinfo {author} {\bibfnamefont {M.~J.}\ \bibnamefont {Steel}}, \bibinfo
  {author} {\bibfnamefont {C.}~\bibnamefont {Martijn~de Sterke}},\ and\
  \bibinfo {author} {\bibfnamefont {R.~C.}\ \bibnamefont {McPhedran}},\
  }\bibfield  {title} {\bibinfo {title} {Spontaneous emission suppression via
  quantum path interference in coupled microcavities},\ }\href@noop {}
  {\bibfield  {journal} {\bibinfo  {journal} {Phys. Rev. Lett.}\ }\textbf
  {\bibinfo {volume} {96}},\ \bibinfo {pages} {103902} (\bibinfo {year}
  {2006})}\BibitemShut {NoStop}%
\bibitem [{\citenamefont {Quesada}\ \emph {et~al.}(2020)\citenamefont
  {Quesada}, \citenamefont {Triginer}, \citenamefont {Vidrighin},\ and\
  \citenamefont {Sipe}}]{twinbeamwaveguide}%
  \BibitemOpen
  \bibfield  {author} {\bibinfo {author} {\bibfnamefont {N.}~\bibnamefont
  {Quesada}}, \bibinfo {author} {\bibfnamefont {G.}~\bibnamefont {Triginer}},
  \bibinfo {author} {\bibfnamefont {M.~D.}\ \bibnamefont {Vidrighin}},\ and\
  \bibinfo {author} {\bibfnamefont {J.~E.}\ \bibnamefont {Sipe}},\ }\bibfield
  {title} {\bibinfo {title} {Theory of high-gain twin-beam generation in
  waveguides: From maxwell's equations to efficient simulation},\ }\href@noop
  {} {\bibfield  {journal} {\bibinfo  {journal} {Phys. Rev. A}\ }\textbf
  {\bibinfo {volume} {102}},\ \bibinfo {pages} {033519} (\bibinfo {year}
  {2020})}\BibitemShut {NoStop}%
\bibitem [{\citenamefont {Fussell}\ and\ \citenamefont
  {Dignam}(2007)}]{engineeringQ}%
  \BibitemOpen
  \bibfield  {author} {\bibinfo {author} {\bibfnamefont {D.~P.}\ \bibnamefont
  {Fussell}}\ and\ \bibinfo {author} {\bibfnamefont {M.~M.}\ \bibnamefont
  {Dignam}},\ }\bibfield  {title} {\bibinfo {title} {Engineering the quality
  factors of coupled-cavity modes in photonic crystal slabs},\ }\href@noop {}
  {\bibfield  {journal} {\bibinfo  {journal} {Appl. Phys. Lett.}\ }\textbf
  {\bibinfo {volume} {90}},\ \bibinfo {pages} {183121} (\bibinfo {year}
  {2007})}\BibitemShut {NoStop}%
\bibitem [{\citenamefont {Breuer}\ and\ \citenamefont
  {Petruccione}(2002)}]{openQsystemsBreuer}%
  \BibitemOpen
  \bibfield  {author} {\bibinfo {author} {\bibfnamefont {H.~P.}\ \bibnamefont
  {Breuer}}\ and\ \bibinfo {author} {\bibfnamefont {F.}~\bibnamefont
  {Petruccione}},\ }\href@noop {} {\emph {\bibinfo {title} {The Theoy of Open
  Quantum Systems}}}\ (\bibinfo  {publisher} {Oxford University Press},\
  \bibinfo {year} {2002})\BibitemShut {NoStop}%
\bibitem [{\citenamefont {Dignam}\ and\ \citenamefont
  {Dezfouli}(2012)}]{photonQuantumDot}%
  \BibitemOpen
  \bibfield  {author} {\bibinfo {author} {\bibfnamefont {M.~M.}\ \bibnamefont
  {Dignam}}\ and\ \bibinfo {author} {\bibfnamefont {M.~K.}\ \bibnamefont
  {Dezfouli}},\ }\bibfield  {title} {\bibinfo {title} {Photon-quantum-dot
  dynamics in coupled-cavity photonic crystal slabs},\ }\href@noop {}
  {\bibfield  {journal} {\bibinfo  {journal} {Phys. Rev. A}\ }\textbf {\bibinfo
  {volume} {85}},\ \bibinfo {pages} {013809} (\bibinfo {year}
  {2012})}\BibitemShut {NoStop}%
\bibitem [{\citenamefont {Garrison}\ and\ \citenamefont
  {Chiao}(2008)}]{quantumopticsGarrison}%
  \BibitemOpen
  \bibfield  {author} {\bibinfo {author} {\bibfnamefont {J.~C.}\ \bibnamefont
  {Garrison}}\ and\ \bibinfo {author} {\bibfnamefont {R.~Y.}\ \bibnamefont
  {Chiao}},\ }\href@noop {} {\emph {\bibinfo {title} {Quantum Optics}}}\
  (\bibinfo  {publisher} {Oxford University Press},\ \bibinfo {year}
  {2008})\BibitemShut {NoStop}%
\bibitem [{\citenamefont {Vendromin}\ and\ \citenamefont
  {Dignam}(2020)}]{onemodeSL}%
  \BibitemOpen
  \bibfield  {author} {\bibinfo {author} {\bibfnamefont {C.}~\bibnamefont
  {Vendromin}}\ and\ \bibinfo {author} {\bibfnamefont {M.~M.}\ \bibnamefont
  {Dignam}},\ }\bibfield  {title} {\bibinfo {title} {Optimization of a lossy
  microring resonator system for the generation of quadrature-squeezed
  states},\ }\href@noop {} {\bibfield  {journal} {\bibinfo  {journal} {Phys.
  Rev. A}\ }\textbf {\bibinfo {volume} {102}},\ \bibinfo {pages} {023705}
  (\bibinfo {year} {2020})}\BibitemShut {NoStop}%
\bibitem [{\citenamefont {Ma}\ and\ \citenamefont
  {Rhodes}(1990)}]{multimodeSqueezeOp}%
  \BibitemOpen
  \bibfield  {author} {\bibinfo {author} {\bibfnamefont {X.}~\bibnamefont
  {Ma}}\ and\ \bibinfo {author} {\bibfnamefont {W.}~\bibnamefont {Rhodes}},\
  }\bibfield  {title} {\bibinfo {title} {Multimode squeeze operators and
  squeezed states},\ }\href@noop {} {\bibfield  {journal} {\bibinfo  {journal}
  {Phys. Rev. A}\ }\textbf {\bibinfo {volume} {41}},\ \bibinfo {pages} {4625}
  (\bibinfo {year} {1990})}\BibitemShut {NoStop}%
\bibitem [{BCH()}]{BCH}%
  \BibitemOpen
  \href@noop {} {}\bibinfo {note} {The formula we use is ${\rm e}^{-A}B{\rm
  e}^A = B - [A,B] + \frac{1}{2!}[A,[A,B]] - \frac{1}{3!}[A,[A,[A,B]]]+\ldots$,
  where $A$ and $B$ are general operators.}\BibitemShut {Stop}%
\bibitem [{\citenamefont {Bunse-Gerstner}\ and\ \citenamefont
  {Gragg}(1988)}]{divideConquerTakagi}%
  \BibitemOpen
  \bibfield  {author} {\bibinfo {author} {\bibfnamefont {A.}~\bibnamefont
  {Bunse-Gerstner}}\ and\ \bibinfo {author} {\bibfnamefont {W.~B.}\
  \bibnamefont {Gragg}},\ }\bibfield  {title} {\bibinfo {title} {Singular value
  decompositions of complex symmetric matrices},\ }\href@noop {} {\bibfield
  {journal} {\bibinfo  {journal} {JCAM}\ }\textbf {\bibinfo {volume} {21}},\
  \bibinfo {pages} {41} (\bibinfo {year} {1988})}\BibitemShut {NoStop}%
\bibitem [{\citenamefont {Masada}\ \emph {et~al.}(2015)\citenamefont {Masada},
  \citenamefont {Miyata}, \citenamefont {Politi}, \citenamefont {Hashimoto},
  \citenamefont {O'Brien},\ and\ \citenamefont {Furusawa}}]{onChipCV}%
  \BibitemOpen
  \bibfield  {author} {\bibinfo {author} {\bibfnamefont {G.}~\bibnamefont
  {Masada}}, \bibinfo {author} {\bibfnamefont {K.}~\bibnamefont {Miyata}},
  \bibinfo {author} {\bibfnamefont {A.}~\bibnamefont {Politi}}, \bibinfo
  {author} {\bibfnamefont {T.}~\bibnamefont {Hashimoto}}, \bibinfo {author}
  {\bibfnamefont {J.~L.}\ \bibnamefont {O'Brien}},\ and\ \bibinfo {author}
  {\bibfnamefont {A.}~\bibnamefont {Furusawa}},\ }\bibfield  {title} {\bibinfo
  {title} {Continuous-variable entanglement on a chip},\ }\href@noop {}
  {\bibfield  {journal} {\bibinfo  {journal} {Nat. Photonics}\ }\textbf
  {\bibinfo {volume} {9}},\ \bibinfo {pages} {316} (\bibinfo {year}
  {2015})}\BibitemShut {NoStop}%
\bibitem [{\citenamefont {Duan}\ \emph {et~al.}(2000)\citenamefont {Duan},
  \citenamefont {Giedke}, \citenamefont {Cirac},\ and\ \citenamefont
  {Zoller}}]{Duan2000InseparabilitySystems}%
  \BibitemOpen
  \bibfield  {author} {\bibinfo {author} {\bibfnamefont {L.-M.}\ \bibnamefont
  {Duan}}, \bibinfo {author} {\bibfnamefont {G.}~\bibnamefont {Giedke}},
  \bibinfo {author} {\bibfnamefont {J.~I.}\ \bibnamefont {Cirac}},\ and\
  \bibinfo {author} {\bibfnamefont {P.}~\bibnamefont {Zoller}},\ }\bibfield
  {title} {\bibinfo {title} {{Inseparability criterion for continuous variable
  systems}},\ }\href@noop {} {\bibfield  {journal} {\bibinfo  {journal} {Phys.
  Rev. Lett.}\ }\textbf {\bibinfo {volume} {84}},\ \bibinfo {pages} {2722}
  (\bibinfo {year} {2000})}\BibitemShut {NoStop}%
\bibitem [{\citenamefont {Simon}(2000)}]{Simon2000Peres-HorodeckiSystems}%
  \BibitemOpen
  \bibfield  {author} {\bibinfo {author} {\bibfnamefont {R.}~\bibnamefont
  {Simon}},\ }\bibfield  {title} {\bibinfo {title} {{Peres-Horodecki
  separability criterion for continuous variable systems}},\ }\href@noop {}
  {\bibfield  {journal} {\bibinfo  {journal} {Phys. Rev. Lett.}\ }\textbf
  {\bibinfo {volume} {84}},\ \bibinfo {pages} {2726} (\bibinfo {year}
  {2000})}\BibitemShut {NoStop}%
\bibitem [{\citenamefont {Dinu}\ \emph {et~al.}(2003)\citenamefont {Dinu},
  \citenamefont {Quochi},\ and\ \citenamefont {Garcia}}]{chi3silicon}%
  \BibitemOpen
  \bibfield  {author} {\bibinfo {author} {\bibfnamefont {M.}~\bibnamefont
  {Dinu}}, \bibinfo {author} {\bibfnamefont {F.}~\bibnamefont {Quochi}},\ and\
  \bibinfo {author} {\bibfnamefont {H.}~\bibnamefont {Garcia}},\ }\bibfield
  {title} {\bibinfo {title} {Third-order nonlinearities in silicon at telecom
  wavelengths},\ }\href@noop {} {\bibfield  {journal} {\bibinfo  {journal}
  {Appl. Phys. Lett.}\ }\textbf {\bibinfo {volume} {82}},\ \bibinfo {pages}
  {2954} (\bibinfo {year} {2003})}\BibitemShut {NoStop}%
\bibitem [{\citenamefont {Zippilli}\ and\ \citenamefont
  {Vitali}(2020)}]{anyclusterstate}%
  \BibitemOpen
  \bibfield  {author} {\bibinfo {author} {\bibfnamefont {S.}~\bibnamefont
  {Zippilli}}\ and\ \bibinfo {author} {\bibfnamefont {D.}~\bibnamefont
  {Vitali}},\ }\bibfield  {title} {\bibinfo {title} {Possibility to generate
  any gaussian cluster state by a multimode squeezing transformation},\
  }\href@noop {} {\bibfield  {journal} {\bibinfo  {journal} {Phys. Rev. A}\
  }\textbf {\bibinfo {volume} {102}},\ \bibinfo {pages} {052424} (\bibinfo
  {year} {2020})}\BibitemShut {NoStop}%
\end{thebibliography}%


\providecommand{\noopsort}[1]{}\providecommand{\singleletter}[1]{#1}%
%

\appendix
\section{Connection between SVD and the Takagi factorization}
\label{SVD+takagi}
As mentioned in the text, the Takagi factorization is a special case of the symmetric SVD, where the diagonal matrix from the Takagi factorization is just a scaled version of the singular values from the SVD. In this section we derive the Takagi factorization from the SVD, and show how the diagonal values are related to the singular values. 

The SVD of the nonlinear parameter $G$ is 
\begin{align}
\label{eq:SVD}
G = U \Sigma W^\dagger,
\end{align}
where $\Sigma$ is a diagonal matrix of real and positive singular values, and $U^\dagger U = 1$ and $W^\dagger W = 1$. To obtain the Takagi factorization we define the diagonal complex matrix 
\begin{align}
\label{eq:lambda_matix}
\Lambda = \Sigma W^\dagger U^{*},
\end{align}
where $U^*$ denotes the complex conjugate of $U$. Therefore the Takagi factorization is
\begin{align}
\label{eq:takagi_appendix}
G &= U \Lambda U^{\rm T},
\end{align}
where the SVD can be recovered by putting Eq. \eqref{eq:lambda_matix} into Eq. \eqref{eq:takagi_appendix}, since $U^* U^{\rm T} = 1$. Therefore, to obtain the Takagi factorization of $G$, one can perform the SVD to get the matrix $U$, and multiply the singular value matrix $\Sigma$ by the diagonal matrix $W^\dagger U^*$.

Now we prove that $ W^\dagger U^*$ in Eq. \eqref{eq:lambda_matix} is diagonal. If it is diagonal, then it must be equal to its transpose, such that $ W^\dagger U^* =  U^\dagger W^*$. Taking the Hermitian conjugate of both sides, we obtain
 \begin{align}
\label{eq:DT=D}
   W^{\rm T} U &= U^{\rm T}W.
\end{align}
Moreover, since $G$ is a symmetric matrix we have $G^{\rm T} = G$. Taking the transpose of Eq. \eqref{eq:SVD} we obtain
 \begin{align}
\label{eq:ST=S}
W^*\Sigma U^{\rm T} &= U\Sigma W^\dagger, \nonumber
\\
   \Sigma U^{\rm T} W &= W^{\rm T}U \Sigma, \nonumber
   \\
   \Sigma U^{\rm T} W  &= U^{\rm T} W \Sigma, \nonumber
   \\
   \Sigma U^{\rm T} W \Sigma ^{-1} &= U^{\rm T} W,
\end{align}
where to go from the second line to the third  line we used Eq. \eqref{eq:DT=D}. The last line of Eq. \eqref{eq:ST=S} is the definition of a diagonalizable matrix, where the diagonal form of $U^{\rm T} W$ is the same as $U^{\rm T} W$. Thus, $U^{\rm T} W$ is diagonal and so is its Hermitian conjugate $W^\dagger U^* $. Therefore we have proved that $\Lambda$ is diagonal. 
\section{Details on the derivation of the coupled differential equations for $\Dot{r}_\mu$, $\dot{\phi}$, and $\dot{n}_m$}
\label{appndxsolvingeqs}
In this section we provide the details on the derivation of the coupled differential equations Eqs. \eqref{eq:rdiff} -\eqref{eq:ndiff}.
\subsection{Equations for $\dot{r}_\mu$ and $\dot{\phi}_\mu$}
\label{appndx1}
 To start, we add Eqs. \eqref{eq:DandE_real} and \eqref{eq:DandE_imag}, to obtain
\begin{align}
    \label{eq:D=-E}
D_{ml} = - E_{ml}.
\end{align}
Putting the expression for $D_{ml}$ in Eq. \eqref{eq:Dcoeff} into Eq. \eqref{eq:D=-E}, we obtain
\begin{align}
\label{eq:r_phi_derive_1}
    \frac{x_mx_l - 1}{2\sqrt{x_mx_l}}\sum_\mu U_{m\mu}U_{l\mu} \left(\dot{r}_\mu+\frac{i\dot{\phi}_\mu}{2}\sinh(2r_\mu) \right){\rm e}^{i\phi_\mu} &= - E_{ml}.
\end{align}
Multiplying both sides by $\sum_{m,l}U^*_{\mu m} U^*_{\mu l}$ and using the orthogonality relation $\sum_m U^*_{\nu m}U_{m\mu} = \delta_{\nu \mu}$, we obtain
\begin{align}
\label{eq:r_phi_derive_2}
  \dot{r}_\mu+\frac{i\dot{\phi}_\mu}{2}\sinh(2r_\mu) &= \sum_{m,l} \frac{2\sqrt{x_mx_l}  E_{ml} U^*_{\mu m}U^*_{\mu l}{\rm e}^{-i\phi_\mu}}{1-x_mx_l }.
\end{align}
Equating the real and imaginary parts of both sides of Eq. \eqref{eq:r_phi_derive_2} gives the following equations for $\Dot{r}_\mu$ and $\dot{\phi}_\mu$ 
\begin{align}
\label{eq:r_phi_derive_3}
  \dot{r}_\mu&= \sum_{m,l} \frac{\sqrt{x_mx_l}  \left(E_{ml} U^*_{\mu m}U^*_{\mu l}{\rm e}^{-i\phi_\mu}+E^*_{ml} U_{m \mu}U_{l \mu }{\rm e}^{i\phi_\mu}\right)}{1-x_mx_l },
  \\
  \label{eq:r_phi_derive_4}
  \dot{\phi}_\mu &= \sum_{m,l} \frac{2\sqrt{x_mx_l}  \left(E_{ml} U^*_{\mu m}U^*_{\mu l}{\rm e}^{-i\phi_\mu}-E^*_{ml} U_{m \mu}U_{l \mu }{\rm e}^{i\phi_\mu}\right)}{i(1- x_mx_l)\sinh(2r_\mu) }.
\end{align}
Now, to simplify Eq. \eqref{eq:r_phi_derive_3} and Eq. \eqref{eq:r_phi_derive_4}, we focus on the first term of the sum. Using the expression for $E_{ml}$ in Eq. \eqref{eq:Ecoeff}, we obtain
\begin{widetext}
\begin{align}
    \label{eq:r_phi_derive_5}
    \sum_{m,l} \frac{\sqrt{x_mx_l}  E_{ml} U^*_{\mu m}U^*_{\mu l}{\rm e}^{-i\phi_\mu}}{1-x_mx_l }&= \sum_{\sigma,\nu}\sum_{m,l}U_{m\sigma}U_{l\nu}U^*_{\mu m}U^*_{\mu l}{\rm e}^{i(\phi_\nu-\phi_\mu)}\cosh(r_\sigma)\sinh(r_\nu)\left(i\Omega_{\sigma \nu}+\frac{1}{2}\Gamma_{\sigma \nu} \frac{1+x_mx_l-2x_m}{1-x_mx_l} \right) \nonumber
    \\
    &- \frac{i}{\hbar}\sum_\nu\sum_{m,l} U_{m\nu}U_{l\nu}U^*_{\mu m}U^*_{\mu l}{\rm e}^{i(\phi_\nu-\phi_\mu)} \left(\alpha^2\lambda_\nu{\rm e}^{-i\phi_\nu} \cosh^2(r_\nu) + \alpha^{*2}\lambda^*_\nu {\rm e}^{i\phi_\nu}\sinh^2(r_\nu)\right) \nonumber
    \\
    &= i\Omega_{\mu\mu}\cosh(r_\mu)\sinh(r_\mu) - \frac{i}{\hbar}\left(\alpha^2\lambda_\mu{\rm e}^{-i\phi_\mu} \cosh^2(r_\mu) + \alpha^{*2}\lambda^*_\mu {\rm e}^{i\phi_\mu}\sinh^2(r_\mu)\right) \nonumber
 \\
    &+\frac{1}{2}\sum_{\sigma,\nu}{\rm e}^{i(\phi_\nu-\phi_\mu)}\Gamma_{\sigma \nu}\cosh(r_\sigma)\sinh(r_\nu) \sum_{m,l}U_{m\sigma}U_{l\nu}U^*_{\mu m}U^*_{\mu l}\frac{1+x_mx_l-2x_m}{1-x_mx_l}. 
\end{align}
\end{widetext}
Using Eq. \eqref{eq:xtrans} we can write Eq. \eqref{eq:r_phi_derive_5} in terms of the thermal photon numbers as
\begin{widetext}
\begin{align}
    \label{eq:r_phi_derive_6}
    \sum_{m,l} \frac{\sqrt{x_mx_l}  E_{ml} U^*_{\mu m}U^*_{\mu l}{\rm e}^{-i\phi_\mu}}{1-x_mx_l }&= i\Omega_{\mu\mu}\cosh(r_\mu)\sinh(r_\mu) - \frac{i}{\hbar}\left(\alpha^2\lambda_\mu{\rm e}^{-i\phi_\mu} \cosh^2(r_\mu) + \alpha^{*2}\lambda^*_\mu {\rm e}^{i\phi_\mu}\sinh^2(r_\mu)\right) \nonumber
 \\
    &-\frac{1}{2}\sum_{\sigma,\nu}{\rm e}^{i(\phi_\nu-\phi_\mu)}\Gamma_{\sigma \nu}\cosh(r_\sigma)\sinh(r_\nu) \sum_{m,l}U_{m\sigma}U_{l\nu}U^*_{\mu m}U^*_{\mu l}\frac{-n_m+n_l+1}{n_m+n_l+1}. 
\end{align}
\end{widetext}
The complex conjugate of Eq. \eqref{eq:r_phi_derive_6} is
\begin{widetext}
\begin{align}
    \label{eq:r_phi_derive_7}
    \sum_{m,l} \frac{\sqrt{x_mx_l}  E^*_{ml} U_{\mu m}U_{\mu l}{\rm e}^{i\phi_\mu}}{1-x_mx_l }&= -i\Omega_{\mu\mu}\cosh(r_\mu)\sinh(r_\mu) + \frac{i}{\hbar}\left(\alpha^2\lambda_\mu{\rm e}^{-i\phi_\mu} \sinh^2(r_\mu) + \alpha^{*2}\lambda^*_\mu {\rm e}^{i\phi_\mu}\cosh^2(r_\mu)\right) \nonumber
 \\
    &-\frac{1}{2}\sum_{\sigma,\nu}{\rm e}^{-i(\phi_\nu-\phi_\mu)}\Gamma^*_{\sigma \nu}\cosh(r_\sigma)\sinh(r_\nu) \sum_{m,l}U^*_{m\sigma}U^*_{l\nu}U_{\mu m}U_{\mu l}\frac{-n_m+n_l+1}{n_m+n_l+1},
\end{align}
\end{widetext}
where we have used the fact that $\Omega^*_{\mu\mu} = \Omega_{\mu \mu}$. Adding Eq. \eqref{eq:r_phi_derive_6} and Eq. \eqref{eq:r_phi_derive_7} together gives the equation for $\dot{r}_\mu$ in Eq. \eqref{eq:rdiff} in the main text. Subtracting Eq. \eqref{eq:r_phi_derive_7} from Eq. \eqref{eq:r_phi_derive_6} and dividing by $\frac{i}{2}\sinh(2r_\mu)$ (see Eq. \eqref{eq:r_phi_derive_4}) gives the equation for $\dot{\phi}_\mu$ in Eq. \eqref{eq:phidiff} in the main text.
\subsection{Equation for $\Dot{n}_m$}
\label{appndx2}
As mentioned in Sec. \ref{differentialequations} we obtain an equation for $\dot{n}_m$ by writing the operators $b^\dagger_mb_m$ and $b^\dagger_m b_l$  found in the expressions $T1$, $T2$, and $T3$ in terms of the Schmidt operator $B^\dagger_\mu B_\nu$, and then force the sum of the coefficients in front of $B^\dagger_\mu B_\nu$ equal to zero. The appropriate term in $T3$ ($i.e.$ Eq. \eqref{eq:term3}) becomes
\begin{align}
    \label{eq:ndiff_derive_1}
    -\sum_m b^\dagger_m b_m \frac{\dot{x}_m}{x_m} = -\sum_{\mu,\nu} \sum_m U^*_{m\mu}U_{m \nu}\frac{\dot{x}_m}{x_m}B^\dagger_\mu B_\nu,
\end{align}
where we have used Eq. \eqref{eq:reversetransformation}. The appropriate terms in $T1$ and $T2$ ($i.e.$ Eq. \eqref{eq:term1_3} and Eq. \eqref{eq:term2_2}) become
\begin{align}
    \label{eq:ndiff_derive_2}
   \sum_{m,l} F_{ml} b^\dagger_m b_l = \sum_{\mu,\nu} \sum_{m,l}  F_{ml} U^*_{m\mu}U_{l \nu}B^\dagger_\mu B_\nu,
\end{align}
and
\begin{align}
    \label{eq:ndiff_derive_3}
    \sum_{m,l} K_{ml} b^\dagger_m b_l = \sum_{\mu,\nu} \sum_{m,l}  K_{ml} U^*_{m\mu}U_{l \nu}B^\dagger_\mu B_\nu.
\end{align}
Since we require that $0=T1+T2+T3$ (see Eq. \eqref{eq:maineq}), the sum of the coefficients multiplying $B^\dagger_\mu B_\nu$ in Eqs. \eqref{eq:ndiff_derive_1} - \eqref{eq:ndiff_derive_3} must be equal to zero
\begin{align}
 \label{eq:ndiff_derive_4}
    \sum_{m,l} U^*_{m\mu}U_{l \nu}\left(F_{ml}+K_{ml} -\frac{\dot{x}_m}{x_m} \right)=0.
\end{align}
Multiplying Eq. \eqref{eq:ndiff_derive_4} by $\sum_{\mu,\nu}U^*_{m \nu}U_{m\mu}$ and using the orthogonality relation $\sum_\mu U^*_{m \mu} U_{l \mu} = \delta_{ml}$, we obtain
\begin{align}
    \label{eq:ndiff_derive_5}
    F_{mm}+K_{mm} -\frac{\dot{x}_m}{x_m}  = 0,
\end{align}
but from Eq. \eqref{eq:Ncoeff} we have that $F_{mm} = 0$. Therefore
\begin{align}
    \label{eq:ndiff_derive_6}
 \frac{\dot{x}_m}{x_m}  = K_{mm}.
\end{align}
Using Eq. \eqref{eq:Gcoeff} in Eq. \eqref{eq:ndiff_derive_6}, we obtain
\begin{align}
 \label{eq:ndiff_derive_7}
 \frac{\dot{x}_m}{1 - x_m}  &=
    \sum_{\mu,\nu}U_{m\mu}U^*_{m\nu}\bigg( -x_m\Gamma_{\mu \nu}\cosh(r_\mu)\cosh(r_\nu)\nonumber
    \\
    &+\Gamma^*_{\mu \nu} {\rm e}^{i(\phi_\mu - \phi_\nu)}\sinh(r_\mu)\sinh(r_\nu)\bigg). 
\end{align}
Expressing Eq. \eqref{eq:ndiff_derive_7} in terms of $n_m$ gives
\begin{align}
 \label{eq:ndiff_derive_8}
 \frac{\dot{n}_m}{1 + n_m}  &=
    \sum_{\mu,\nu}U_{m\mu}U^*_{m\nu}\bigg( -\frac{n_m}{1+n_m}\Gamma_{\mu \nu}\cosh(r_\mu)\cosh(r_\nu)\nonumber
    \\
    &+\Gamma^*_{\mu \nu} {\rm e}^{i(\phi_\mu - \phi_\nu)}\sinh(r_\mu)\sinh(r_\nu)\bigg),
\end{align}
which is the same as Eq. \eqref{eq:ndiff} in the text.
\section{Solving the coupled equations, Eqs. \eqref{eq:rdiff} - \eqref{eq:ndiff}}
\label{solving}
In this section we discuss how to solve  Eqs. \eqref{eq:rdiff} - \eqref{eq:ndiff} for a system that initially starts in the \textit{vacuum state}. At time $t=0$ the vacuum state is defined by
\begin{align}
\label{eq:r_t=0}
    r_\mu(0) &= 0,
    \\
    \label{eq:n_t=0}
    n_m(0)&=0,
\end{align}
for all $\mu$ and $m$. Putting these initial conditions into Eq. \eqref{eq:rdiff} we obtain
\begin{align}
    \label{eq:rdiff_t=0}
    \Dot{r}_\mu(0)&=\frac{2\left|\alpha(0)\right|^2\left|\lambda_\mu\right|}{\hbar}\sin\left(-\phi_\mu(0) + \theta_\mu\right),
\end{align}
where we define 
\begin{align}
    \alpha(t)^2 &= |\alpha(t)|^2\exp(2i\omega_P t),
\end{align}
 and 
 \begin{align}
     \lambda_\mu = |\lambda_\mu|\exp(i\theta_\mu),
 \end{align}
 where $\theta_\mu$ is a real number. We choose the initial squeezing phase, $\phi_\mu(0)$, to be
\begin{align}
    \label{eq:phi_t=0}
    \phi_\mu(0) = \theta_\mu - \frac{\pi}{2},
\end{align}
such that is maximizes the squeezing amplitude at the next time-step, $r_\mu(\Delta t)$:
\begin{align}
\label{eq:r_t=dt}
    r_\mu(\Delta t) = \frac{2\left|\alpha(0)\right|^2\left|\lambda_\mu\right|}{\hbar} \Delta t + O\left(\left(\Delta t\right)^2\right).
\end{align}
Now, let us move on to the equation for the squeezing phase,  Eq. \eqref{eq:phidiff}. It is easily shown that using Eqs. \eqref{eq:r_t=0}, \eqref{eq:n_t=0}, and \eqref{eq:phi_t=0}
will result in the second and third terms in Eq. \eqref{eq:phidiff} being indeterminate (0/0). Thus, at $t=0$, we write this equation as
\begin{align}
    \label{eq:phidiff_t=0}
    \Dot{\phi}_\mu(0)&=2\Omega_{\mu\mu}-\zeta_\mu,
\end{align}
where we let $\zeta_\mu$ be the indeterminate form. We are unable to solve these equations unless we define $\zeta_\mu$. We define $\zeta_\mu$ by requiring that the derivatives of the squeezing phase at $t=\Delta t$ and $t=0$ are the same, 
\begin{align}
\label{eq:ddphi=0}
    \dot{\phi}_\mu(\Delta t)=\Dot{\phi}_\mu(0),
\end{align}
such that initially the squeezing phase is a linear function of time. Putting Eqs. \eqref{eq:r_t=0}, \eqref{eq:n_t=0}, \eqref{eq:phi_t=0}, and \eqref{eq:r_t=dt} into Eq. \eqref{eq:phidiff} and using the fact that $n_m(\Delta t) = 0$ (which can be proven by writing Eq. \eqref{eq:ndiff} as a difference equation and using the initial conditions) it can be shown that
\begin{align}
\label{eq:dphi_t=dt}
    \dot{\phi}_\mu(\Delta t)=2\omega_P+\zeta_\mu.
\end{align}
Using Eq. \eqref{eq:dphi_t=dt} in Eq. \eqref{eq:ddphi=0}, the indeterminate form $\zeta_\mu$ is defined as
\begin{align}
\label{eq:indeterminate}
    \zeta_\mu = \Omega_{\mu \mu} - \omega_P.
\end{align}

To solve  Eqs. \eqref{eq:rdiff} - \eqref{eq:ndiff} we use MATLAB's ode45 function, that is based on a Runge-Kutta method.  The initial conditions that we use for the squeezing amplitudes, thermal photon numbers, and squeezing phases are in Eqs. \eqref{eq:r_t=0}, \eqref{eq:n_t=0}, and \eqref{eq:phi_t=0}. We have to write an additional condition in the code that imposes the condition that at $t=0$  the derivatives of the squeezing phases are equal to $\dot{\phi}_\mu(0) = \Omega_{\mu\mu} + \omega_P$, otherwise the program will return a division-by-zero error (as discussed above). The solution is sensitive to the initial squeezing phases $\phi_\mu(0)$. It is crucial that they are set to precisely the values given in Eq. \eqref{eq:phi_t=0} in order to obtain the results we present in Sec. \ref{results}. We find, however, that the initial value of the derivative of the phase $\dot{\phi}_\mu(0)$ has little impact on the final solution, since it quickly settles to the correct value, given by $\Omega_{\mu\mu} + \omega_P$.

\end{document}